\documentclass[sigplan,screen,nonacm]{acmart}
\usepackage{moreverb}
\usepackage{subcaption}
\usepackage{catchfile}
\usepackage{algorithm}
\usepackage{algpseudocode}
\usepackage{graphicx}
\usepackage{textcomp}
\usepackage{xcolor}
\usepackage{listings}
\usepackage{hyperref}
\usepackage{makecell}
\usepackage{multirow}
\usepackage{tikz}

\newcommand\BibTeX{{\rmfamily B\kern-.05em \textsc{i\kern-.025em b}\kern-.08em
T\kern-.1667em\lower.7ex\hbox{E}\kern-.125emX}}

\DeclareMathAlphabet{\pazocal}{OMS}{zplm}{m}{n}
\newcommand{\unif}{\pazocal{U}}

\setcopyright{none}

\begin{document}

\title{Task-based preemptive scheduling on FPGAs leveraging partial reconfiguration}

\author{Gabriel Rodriguez-Canal}
%%\authornote{Both authors contributed equally to this research.}
\email{gabriel.rodcanal@ed.ac.uk}
\affiliation{%
  \institution{EPCC at the University of Edinburgh}
  \streetaddress{Bayes Centre, 47 Potterrow}
  \city{Edinburgh}
  \country{United Kingdom}
}

\author{Nick Brown}
\orcid{0000-0003-2925-7275}
\affiliation{%
  \institution{EPCC at the University of Edinburgh}
  \streetaddress{47 Potterrow}
  \city{Edinburgh}
  \country{UK}
}

\author{Yuri Torres}
\affiliation{%
  \institution{Escuela de Ingeniería Informática}
  \streetaddress{Universidad de Valladolid}
  \city{Valladolid}
  \country{Spain}
}

\author{Arturo Gonzalez-Escribano}
\affiliation{%
  \institution{Escuela de Ingeniería Informática}
  \streetaddress{Universidad de Valladolid}
  \city{Valladolid}
  \country{Spain}
}

\begin{abstract}
FPGAs are an attractive type of accelerator for all-purpose HPC computing systems due to the possibility of deploying tailored hardware on demand. However, the common tools for programming and operating FPGAs are still complex to use, especially in scenarios where diverse types of tasks should be dynamically executed. In this work we present a programming abstraction with a simple interface that internally  leverages High-Level Synthesis, Dynamic Partial Reconfiguration and synchronisation mechanisms to use an FPGA as a multi-tasking server with preemptive scheduling and priority queues. This leads to an improved use of the FPGA resources, allowing the execution of several different kernels concurrently and deploying  the most urgent ones as fast as possible.

The results of our experimental study show that our approach incurs only a 10$\pm$5\% overhead in the worst case  when using two reconfigurable regions, whilst providing a significant performance improvement of at least 24$\pm$21\% over the traditional full reconfiguration approach.
\end{abstract}


\maketitle

\section{Introduction}

The end of Moore's law %\cite{moore1998cramming} 
and loss of Dennard's scaling %\cite{dennard2007design} 
has motivated the search for alternative ways of improving the performance of high-performance computational systems. As a result, heterogeneous systems, primarily composed of CPUs and GPUs have become commonplace in modern High Performance Computing (HPC) machines \cite{top500}. However, these architectures are not ideally suited for all codes, and it has been found that when HPC applications are bound by aspects other than compute, for instance memory bound codes, moving to a dataflow style and exploiting the specialisation of FPGAs can be beneficial \cite{brown2019s,brown2020exploring,brown2021porting}. Nonetheless, FPGAs have not yet been adopted by any of the large supercomputers, which is due to the challenges associated with both their programmability and flexibility. The former has been partially addressed by High Level Synthesis (HLS) tooling, enabling the programmer to write their code in C or C++. However the latter has been less explored. This is because the entire FPGA is often stalled during fabric reconfiguration which means that dynamic scheduling and preemptive execution of workloads is uncommon. 

In this paper we propose a programming abstraction to enable the leveraging of an FPGA
as a multi-tasking computational resource with preemptive scheduling and priority queues. We utilise both the traditional \emph{full reconfiguration} mechanism that halts and wipes the entire FPGA, and the \emph{Dynamic Partial Reconfiguration} (DPR) mechanism, reconfiguring several partial areas of the FPGA independently enabling these to execute different kernels. The work described in this paper hides the complex and low-level details associated with using dynamic partial reconfiguration and synchronisation mechanisms to support on-the-fly instantiation, stopping and resumption of kernels on specific parts of the FPGA fabric whilst the rest of the chip continues executing other workloads independently. In our approach, tasks are programmed as OpenCL kernels and managed with the Controller model \cite{moreton2017supporting,moreton2018controllers,rodriguez2021efficient}, which is a heterogeneous programming model implemented as a C99 library of functions. Controller aims to efficiently manage different types of devices with a portable interface and in this paper we describe an extension to support multiple kernels and preemption on dynamic partial reconfiguration capable FPGA systems. Furthermore, the scheduler that we have developed supports First-Come-First-Served (FCFS) preemptive scheduler with priority queues, significantly increasing the flexibility of Controller and how it maps tasks to the FPGA. It should be noted that we largely use the term \emph{task} and \emph{kernel} interchangeably in this paper.

This work extends a previous prototype presented in~\cite{HETEROPAR2022}, significantly enhancing the scheduler to provide improved policies and lower overhead due to a simplification of the synchronisation mechanisms. Furthermore, we have also decoupled our two techniques for dynamic partial reconfiguration and kernel interruption/resumption, allowing for preemption when using full FPGA reconfiguration. 
%In this work we improve our scheduler presented in [CITE HETEROPAR2022], as described in Section \ref{sect:dpr_scheduler}}
%support for FR or DPR alternatives,
%a comparison with the traditional Full Reconfiguration mechanism, 
 %Finally, we provide a framework for the generation of such DPR system and the corresponding partial bitstreams for all the permutations of the kernels provided by the user.
The rest of the paper is organised as follows: Section \ref{sect:rel_work} describes the background to this work, surveying FPGAs and the Controller task-based model, before exploring related activities which tackle flexible execution of kernels on FPGAs. This is then followed by Section~\ref{sec:approach} where we present the techniques and extensions that have been developed as part of this work, both on the on-chip FPGA infrastructure side and also on the host side including integration with Controller and development of a FCFS scheduler. Section \ref{sec:abstractions} then describes the programming level abstractions provided to the user to enable convenient FPGA kernel state storage and loading, supporting the preemption of kernels on reconfigurable architectures. Section \ref{sec:experiments} reports the results of an
experimental study where we evaluate our approach, and lastly Section \ref{sect:conclusions} draws conclusions and discusses further work.

The novel contributions of this paper are:
\begin{itemize}
\item An approach which provides the ability to undertake low overhead, fine grained, task swapping on an FPGA which delivers the flexibility of task-based models to reconfigurable architectures.
\item Support for the preemption of FPGA kernels, where higher priority kernels can interrupt running tasks which will then be restarted, based on their current computed state, at a later time.
\item An experimental study to explore the performance properties of using dynamic partial reconfiguration and FPGA kernel preemption.
\end{itemize}

\section{Background and related work} 
\label{sect:rel_work}

\subsection{Field Programmable Gate Arrays (FPGAs)}

Field Programmable Gate Arrays (FPGAs) are a type of reconfigurable architecture where the electronics of the chip can be configured at runtime to undertake specific activities. This ability to tailor the electronics of the chip to the workload in question is potentially beneficial when compared against more general purpose architectures, such as CPUs or GPUs, for high performance workloads for a number of reasons. Firstly, codes which are memory or microarchitecture bound on general purpose architectures can benefit from this specialisation because the electronics executes the algorithm directly, without black-box layers comprising the general purpose micro-architecture and associated assumptions \cite{brown2020exploring}. Secondly, removing the need to support aspects within the CPU's microarchitecture such as branch prediction, precise exceptions and register renaming, which can add considerable complexity, can result in reduced power draw and the ability to leverage more of the chip for computation. Lastly, it has been observed that FPGAs often deliver good strong scaling performance, especially with smaller local problem sizes due to CPU or GPU having to \emph{ramp up} the clock frequency which the FPGA does not need to do. \cite{menzel2021strong}

A tradition way of describing FPGAs is as containing large amounts of reconfigurable logic which are sitting within a sea of reconfigurable interconnect. However, whilst this is correct, it oversimplifies the technology because, in addition to the reconfigurable logic provided via Look Up Tables (LUTs), modern FPGAs also contain fast on-chip memories, such as Block RAM (BRAM), and Digital Signal Processing (DSP) slices that can be used to undertake arithmetic operations. FPGAs are configured via a binary bitstream, which will set up the constituent parts of the FPGA as appropriate. These bitstreams are built via a vendor's tooling, and in this paper we focus on AMD Xilinx FPGAs where their Vivado tool is used to design and implement the FPGA logic. Whilst full reconfiguration of the entire FPGA is most common, more recently Dynamic Partial Reconfiguration (DPR) has become possible, where only specific parts of the FPGA are reconfigured whilst other parts of the chip are unaffected and can continue to run independently.

Historically, FPGAs were programmed via esoteric Hardware Description Languages (HDL), such as Verilog and VHDL. However more recently High Level Synthesis (HLS), which enables programmers to write code in C or C++ and for this to be synthesised down to the HDL level, have become more popular. Combined with the very significant investment being made by vendors in their FPGA software ecosystems, writing codes for FPGAs is increasingly becoming much more a question of software development rather than hardware design.

In this paper we undertake our experiments upon a Zynq-7020 FPGA mounted on a Pynq-Z2 board. The Zynq is a popular type of FPGA which combines the CPU (known as the Programmable Subsystem) with the configurable fabric (known as the Programmable Logic). The Zynq-7020 contains a dual-core ARM Cortex-A9 which is running Linux, with the reconfigurable programmable logic partition containing 53,200 LUTs, 4.9 Mb of BlockRAM (BRAM), and 220 DSP slices. The host board contains 512MB of DRAM which is accessible to both the CPU and reconfigurable logic, with the CPU driving the reconfiguration of the FPGA based upon a user's bitstream.

\subsection{The task-based programming model and Controller}
In the task-based programming model programmers decompose their problem into distinct tasks which, apart from consuming inputs and generating outputs, are usually self contained. Tasks become eligible to run when all their inputs have been generated from preceding tasks, and are then scheduled for execution on the most appropriate compute resource. The output(s) of a task will then be used as inputs by subsequent tasks in the overall computation.
It has been argued by the community that writing parallel codes which follow the task based model is advantageous for numerous reasons. Firstly the model promotes asynchrony, where decomposing a problem into many small, independent, tasks can significantly reduce the amount of synchronisation that would be imposed by other approaches. Secondly, the scheduling of tasks can often be done in a manner that maps a task to its most appropriate compute resource, for instance to a CPU or GPU, and potentially helps hide the complexities of mixing architectures from the perspective of the programmer. Thirdly, task-based programming can often assist in providing improved levels of resilience, where if a compute resource becomes unavailable then tasks are simply rescheduled elsewhere.

\begin{figure*}[htb]
    \centering
    \includegraphics[width=0.7\textwidth]{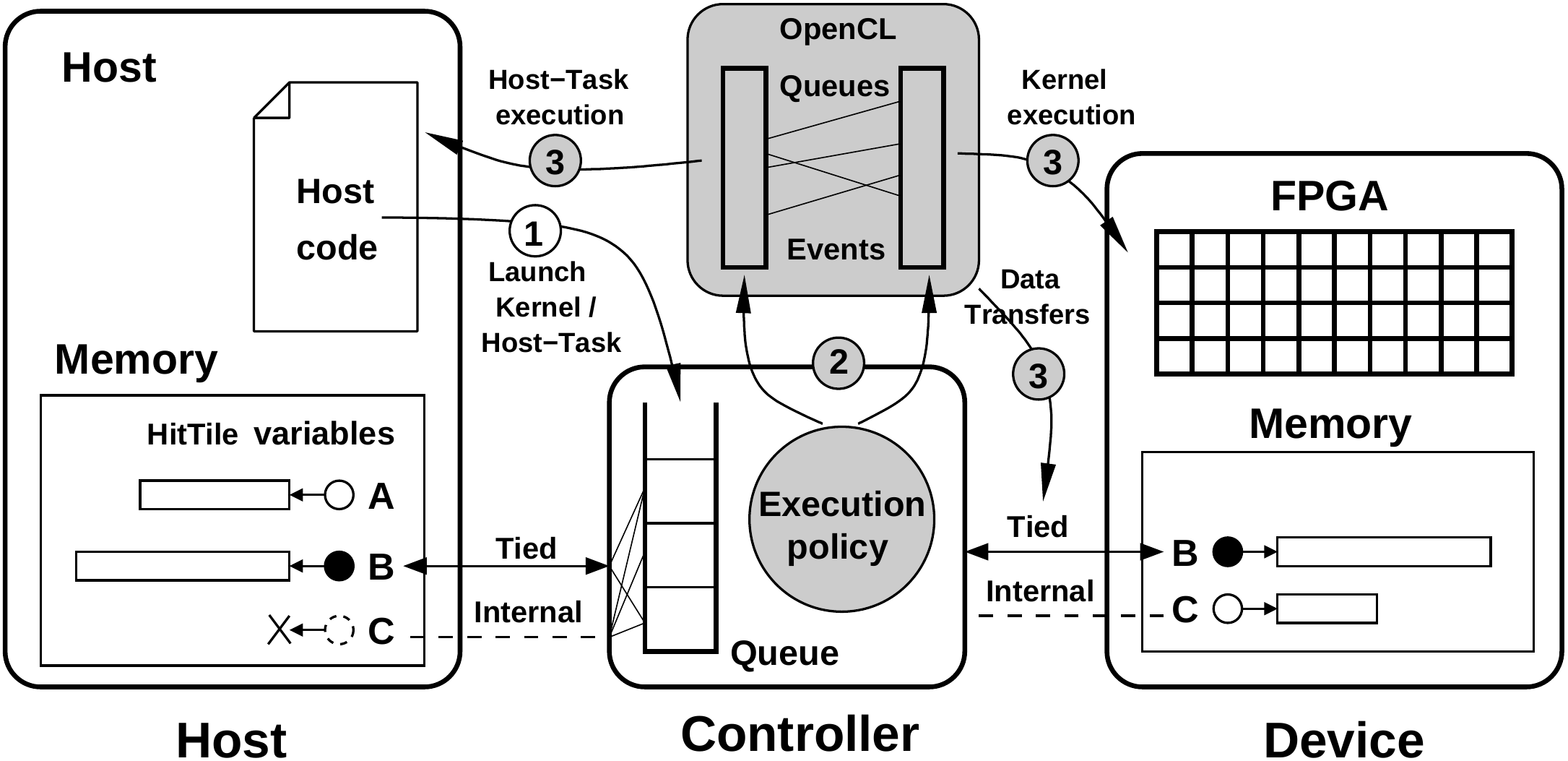}
    \caption{The Controller programming model, generic FPGA backend. Extracted from \cite{rodriguez2021efficient}.}
    \label{fig:controller}
\end{figure*}

There have been many implementations of the task-based model in HPC programming technologies, including OpenMP tasks \cite{ayguade2008design}, OmpSs \cite{bueno2012productive}, and StarPU \cite{augonnet2009starpu}. Controller \cite{moreton2017supporting,moreton2018controllers} is another such example and aims to provide a heterogeneous task-based parallel programming model implemented as a C99 library. The framework provides an abstraction for programming using different types of devices, such as sets of CPU-cores, GPUs, and FPGAs. As illustrated in Figure~\ref{fig:controller}, in Controller the model is based around the \emph{Controller} entity. Each of these is associated with a particular device and manages the execution on, and data-transfers to or from, that device. Programmer's code is executed on the main thread, using the Controller high-level API to enqueue computation tasks for devices. Each Controller entity has its own thread that dequeues and launches the execution of the kernels associated with the tasks.

The Controller runtime resolves data dependencies between tasks, performing data transfers in a manner that is transparent to the end programmer. Requests for kernel execution and data-transfer are stored in three internal queues of the device driver, one controlling the execution order of kernels, one for host-to-device transfer, and one for device-to-host transfers. This design has been demonstrated to provide fast control operation and an efficient overlapping of computation and data transfers where possible \cite{moreton2018controllers}. Portability across architectures is achieved using bespoke runtime backends for different device technologies, such as CUDA or OpenCL, to implement the calls which manage the low-level device queues and events. Furthermore, Controller is agnostic of the language chosen by the programmer for writing their kernel code. Most commonly users leverage OpenCL as this targets the widest range of devices, however more specialised kernels can be written in the native programming models of the accelerator such as CUDA for Nvidia GPUs.

In \cite{rodriguez2021efficient} the authors provided support for Controller to leverage Intel Stratix-10 FPGAs as a target device. However, a major limitation was that full reconfiguration of the entire FPGA was required when mapping a task to the device, meaning that only one kernel could execute on the FPGA at any given time and as such it was common for the compute resource to be significantly under-utilised. The key challenge here was that there was a miss match between the scheduling flexibility commonly provided by task-based models, and the limitations imposed by having to reconfigure the entire FPGA via full reconfiguration.

\subsection{Existing task-based models targeting FPGAs}

Technologies including oneAPI \cite{oneAPI}, OmpSs \cite{bosch2018application}, EngineCL \cite{nozal2020enginecl}, Kokkos \cite{trott2021kokkos}, OpenCL \cite{munshi2009opencl}, and OpenACC \cite{openacc} have been developed or enhanced to support FPGAs. Using these technologies it is possible, although some with a greater level of ease than others, to mix execution across different architectures. 

However, despite improving general programmability by supporting a common host-side programming abstraction, these approaches all fail to fully exploit the full potential of FPGAs. This is mainly a flexibility issue, where these frameworks lack the support do independently swap in and out tasks of varying sizes onto an FPGA accelerator whilst other tasks are running and remain unaffected. Furthermore these frameworks do not support preemption of kernels on the FPGA which, whilst it is secondary in importance to the issue of flexibility, still provides benefit for some workloads.

%The FPGA is programmed with a full bitstream that contains the kernels that will be run during the program execution in a non-preemptive way. % ARTURO: Avoid the term migration on this paper, it is a contribution for the journal paper
%and without the possibility of task migration. 
%Consequently, unlike the CPU or GPU, it is not possible to flexibly multiplex the FPGA between tasks, resulting in a more limited usage mode of the architecture and less opportunity for uptake in supercomputing.

The authors of~\cite{vaishnav2019heterogeneous} tried to address these limitations with their task-based library that targets System on-a Chip (SoC) deployment based on OpenCL using dynamic partial reconfiguration. They support kernel preemption by enabling checkpointing at the end of each OpenCL workgroup, and whilst this is a natural consistency point in the OpenCL model, the coarse-grained nature of the approach limits scheduling flexibility. For example, tasks of higher priority may need to wait until a previous workgroup with lower priority tasks finishes. Moreover, the user must write their kernel interfaces in a manner that are conformant to the interfaces of the FPGA's reconfigurable regions, % ARTURO ???of their design,
causing a conflict between the high-level OpenCL description and the management of the lower-level on-chip infrastructure, which increases the overall development complexity. It is important that this aspect is abstracted from the programmer because of the expertise required on behalf of the programmer to get this right.
Furthermore, the authors use the ZUCL framework \cite{pham2018zucl} to support their approach, which is a predefined static system using the entirety of the FPGA regardless of the number of kernels that are actually being executed by the application.

FPGA support provided within OmpSs \cite{bacardit2021fpga} delivers fast checkpointing on the FPGA via on-chip memory and high-level kernel code annotations. However, their work leverages this mechanism only for fault tolerance and does not consider or support preemptive scheduling.

\section{Approach to support preemptive scheduling on FPGAs}
\label{sec:approach}

In contrast to previous techniques, our approach enables one to leverage the FPGA in a flexible manner through the use of a dynamic partial reconfiguration capable design for each number of reconfigurable regions. This approach avoids potential decreases in the final design clock frequency which can occur because of unnecessarily high occupation, leading to missed opportunities for optimisation in the routing phase of the tooling. To achieve this we focus on two areas; the on-chip FPGA infrastructure and the integration on the host-side of the Controller runtime.

%\subsection{Architecture of our approach} \label{sect:architecture}
\subsection{On-chip infrastructure} \label{sect:architecture}
\begin{figure*}
\begin{center}
    %\includesvg[width=\textwidth, inkscapelatex=false]{img/system_architecture}
    \includegraphics[width=0.7\textwidth]{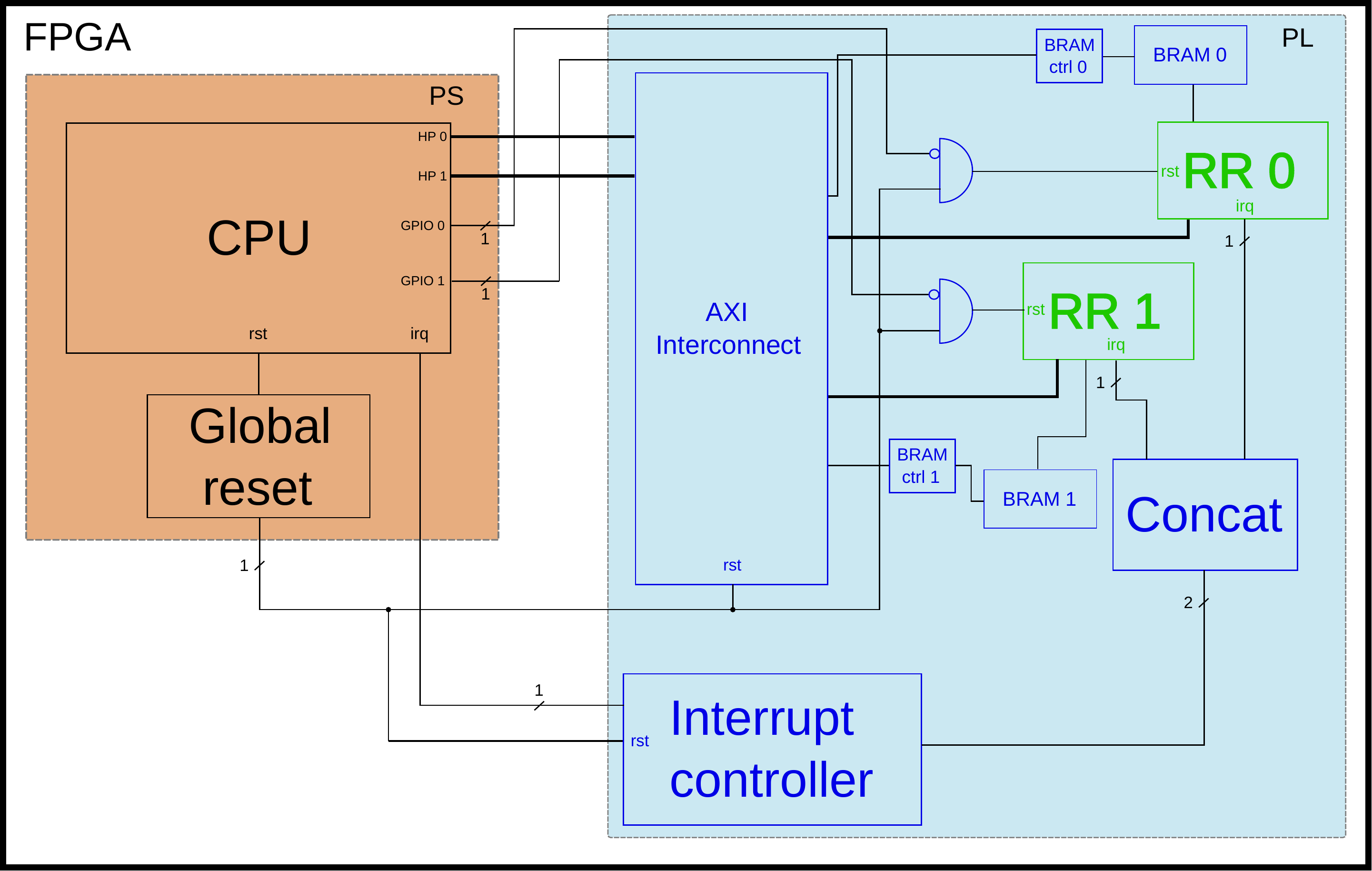}
    \caption{Overview architecture of our shell which supports dynamic partial reconfiguration and preemption of tasks. In this illustration there are two Reconfigurable Regions (RRs) into which kernels are dynamically swapped in and out which is independent of the other region.}
    \label{fig:architecture}
\end{center}
\end{figure*}

Figure~\ref{fig:architecture} illustrates the architecture of the static part of the on-chip infrastructure in our approach. This is known as the \emph{shell}, and is deployed to the FPGA to provide infrastructure and management capabilities such as configuration of the reconfigurable regions and the ability to share memory between the CPU and the reconfigurable logic. The blue box in Figure~\ref{fig:architecture} represents the Programmable Logic (PL) part of the Zynq FPGA, where each constituent box is known as an \emph{IP block}.

The user's tasks, known as kernels at the hardware level, are \emph{dropped into} one of two reconfigurable regions (denoted RR in Figure~\ref{fig:architecture}), although it should be noted that our approach and shell is scalable to any number of regions. Our approach targets HLS kernels that are generated by Xilinx's HLS tooling and using a single master memory interface (known as AXI4-Master) to connect to the board's DRAM memory. Furthermore, a slave interface (known as an AXI4-Slave) is used for the HLS kernel's control ports and this is standard with HLS. It should be noted that, whilst these interfaces are fixed in the design, the user need not use HLS if they do not wish, any type of IP block can be placed into the reconfigurable regions as long as it conforms to the interfaces. Whilst this interface configuration is fairly standard for HLS IP blocks, it is possible to regenerate the shell with a different number and type of interface(s) as required. 

The interrupt controller registers interrupts generated by the reconfigurable regions upon completion, and therefore the CPU can detect when kernels have finished execution. To support preemption of kernels, the shell must be able to interrupt a kernel, saving its context and state, to later resume it. The shell features two on-chip BRAM memory banks (one per reconfigurable region) in which is stored the interrupted kernel's state at arbitrary intervals whose frequency is defined by the user. BRAM memory is used since its speed and closeness to the reconfigurable regions results in very low latency, minimising the overhead of the context saving operation. These BRAM banks are also connected to a BRAM controller which enables access to this memory from the CPU, supporting overall book-keeping of the kernel's state when kernels are being swapped in and out by the scheduler controlling execution from the host.

%Our approach increases complexity around resetting the state of the FPGA, where the ability to 
Our approach increases the complexity of undertaking a reset, as it must be possible to reset both the entire FPGA and individual reconfigurable regions. The former is achieved via the shell's \emph{global reset} (see Figure~\ref{fig:architecture}). The latter is supported by specific reset functionality for each reconfigurable region implemented using the GPIO ports of the CPU. Because, by default, reset is active low for HLS kernels the GPIO signal is negated and then the logical \emph{and} operation is applied with the global reset signal. The application of the reset signal is asynchronous which means that the kernel might be interrupted unpredictably, this is handled by the software abstractions that are described in Section~\ref{sect:abst_preemption} to ensure that a task can be resumed later from a consistent state.

%In order to undertake partial reconfiguration all the interfaces in the RR must be decoupled from the rest of the shell beforehand to prevent corruption of the logic or an undefined state \cite{dfx}. To achieve this all signals and buses connected to the RRs are filtered through a DFX Decoupler, which acts as a multiplexer setting the signals to a constant value in the interfaces when in decoupled mode and otherwise passes the data across unhindered.

This shell design is provided in \emph{netlist} form and packaged as a Tcl script, which is the standard scripting language for Xilinx tools, with the reconfigurable regions instantiated as black boxes. Consequently, to generate the Vivado hardware design one executes the Tcl script with, as an argument, the number of reconfigurable regions that are required. This step is not intended to be completed by the programmer, but instead either by a more FPGA experienced systems team member who will then build and deploy the shell onto the FPGA, or provided pre-built.  

\subsection{Integration into the Controller framework} \label{sec:controller_integrate}
A new Controller backend has been written to interact with the shell design that was described in Section \ref{sect:architecture} which targets the Zynq-7020 FPGA and Pynq-Z2. To undertake low level communication with the FPGA, we use the Pynq C API \cite{pynq-c-api} which exposes rudimentary functionalities, such as the loading of full and partial bitstreams, the interaction with design IP such as interrupt controllers or DMA engines through memory mapping, and host-device shared memory. Building on the C Pynq API means that this work is compatible with any other FPGA from the Zynq-7000 family with little modification required.

Each reconfigurable region is treated by our Controller backend as an independent accelerator, ensuring that tasks can be executed across these regions in parallel. To achieve this we replicate Controller's device specific queue for each reconfigurable region, and each instance is managed by a separate thread. A request to reconfigure a region is implemented as an internal task, queued and executed like any other task. This approach simplifies the backend structure and enforces the scheduling of a reconfiguration request before the associated task execution on the fabric. 

A complication is that it is only possible to reconfigure one region at a time because the Zynq only provides a single Internal Configuration Access Port (ICAP)~\cite{dfx} which is actually used for the partial reconfiguration. Consequently a lock is required to enforce synchronisation between reconfiguration requests in the Controller queues. The Zynq-7000 FPGA family architecture supports shared memory which can be accessed by both the FPGA fabric and host CPU. This memory can be both the fast on-chip BRAM and also the slower off-chip DRAM, and because both the FPGA fabric and CPU can access these spaced directly it means that data-movement operations can be implemented as zero-copy.

Furthermore, the Controller backend leverages Userspace I/O (UIO) to interact with the shell's interrupt controller for detecting interrupts raised by the reconfigurable regions that, as described in Section \ref{sect:architecture}, indicate kernel termination. To avoid constantly polling for these interrupts we use the \emph{select()} system call to activate the manager CPU thread when an interrupt is received. The Controller backend then queries the interrupt controller to determine which region raised the interrupt.

\subsection{The Controller FPGA scheduler} \label{sect:dpr_scheduler}
Building atop the Controller backend described in Section \ref{sec:controller_integrate}, we then used the ability to undertake dynamic partial reconfiguration to build a First Come First Served (FCFS) scheduler for our tasks, which provides the ability to specify task priorities and undertake preemption. 

In order to test the scheduler we simulate scenarios where both the time of the next task arrival and the task parameters are randomly generated. This is undertaken by pre-generating a sequence of tasks, ordered by a random arrival time, where each task has a random priority, a randomly chosen kernel code to execute (from a given set), and random arguments. Our scheduler is modular such that the main scheduling functionality including the management of the queues and service of tasks is separate from our test scenario, which generates random tasks. Consequently it is possible to use this schedule in a test or production mode, and furthermore possible to fairly easily extend or adapt the underlying scheduling logic and to manage any number of reconfigurable regions. 

The scheduler presented in \cite{HETEROPAR2022} was extended to support the preemption of tasks using full reconfiguration. Under this full reconfiguration regime the entire FPGA is halted until reconfiguration has completed, and therefore the state of running tasks must be saved in advance. As a consequence, kernel swap operations are mutually exclusive with kernel tasks, unlike with the dynamic partial reconfiguration approach where reconfiguration latencies are hidden by computation. Upon termination of full reconfiguration the state of the interrupted tasks is restored. 

The main loop of our scheduler is presented in Algorithm \ref{alg:man_loop_scheduler}. It consists of an infinite loop that serves one task per iteration and terminates when all of them have completed. At the beginning of each iteration, the function \emph{waitForInterrupt()} will wait for either one of two events: an interrupt or a timeout. An interrupt is received when a kernel has finished its execution whilst a timeout happens when a new task has arrived. This guarantees a task waits efficiently for a task through the \emph{select()} system call. Line 3 in Algorithm \ref{alg:man_loop_scheduler} tests for the termination condition and, if no such condition is present then tasks are served either as they arrive, between lines 6 and 8, or from the queue whenever a task finishes execution between lines 10 to 14. It can be seen that, whenever the arrival of tasks ceases, then tasks are drawn from the queue and served to the available reconfigurable regions. 
Therefore, as long as there are enough tasks in the queue, the reconfigurable regions will be assigned tasks. This was not ensured in the previous version of the scheduler that was presented in \cite{HETEROPAR2022} which, although it kept the regions at full occupation in many scenarios, was not guaranteed for all mixes of tasks. Finally, line 17 updates the timeout for the next arriving task.

Algorithm \ref{alg:swap_function} presents the swap mechanism of our approach, where it can be seen that full reconfiguration encompasses more steps and book-keeping than dynamic partial reconfiguration. Whenever partial reconfiguration is chosen, the partial bitstream corresponding to the incoming kernel is selected at line 2 and then the scheduler instructs the Controller runtime to perform the reconfiguration at line 3. Otherwise, for full reconfiguration, all the running kernels in the reconfigurable region other than the one requested are preempted to save their state. This step is fundamental to prevent restarting kernels that had already computed work. A simpler approach would adding a synchronisation point for all the kernels before the reconfiguration, but this would add to the overhead of the reconfiguration latency, proportional to the size of the FPGA. Line 10 selects the bitstream that includes the incoming kernel in the requested reconfigurable region, leaving the other kernels unchanged. Line 11 then performs the same task as line 3 and, after the reconfiguration, line 12 sets up the hardware design for operation. This includes setting up and mapping to memory the reconfigurable region's interfaces and controllers, such as for interrupts and BRAM. Finally, between lines 13 and 18, the preempted kernels are restored to resume execution.  

\begin{algorithm}

    \caption{Main loop of the scheduler.}
    \label{alg:man_loop_scheduler}
    \begin{algorithmic}[1]
        %\Comment{Test}
        \While{$true$}
            \State {$waitForInterrupt(timeout)$}
            \If {$hasFinished(N, tasks\_to\_arrive$)}
                \State {$break$}
            \EndIf
            
            \If {$tasks\_to\_arrive \wedge timeout = 0$}
                \State $task \gets getArrivedTask()$
                \State $serveTask(task)$
            \Else
                \ForAll {$r \in R$}
                    \If {$isFree(r)$}
                        \State $task \gets getTaskFromQueue()$
                        \State $serveTask(task)$
                    \EndIf
                \EndFor
            \EndIf
            \State $updateTimeout(timeout)$
        \EndWhile
    \end{algorithmic}
\end{algorithm}

\begin{algorithm}
    \caption{Swap function}
    \label{alg:swap_function}
    \begin{algorithmic}[1]
        \If {$partial$}
            \State $partial\_bitstream \gets get\_partial\_bitstream(incoming\_task)$
            \State $doSwap(partial\_bitstream)$
        \Else
            \ForAll {$r \in R$}
                \If {$running(r) \wedge r \neq r_{swap}$}
                    \State $evictKernel(r)$
                \EndIf
            \EndFor

            \State $full\_bitstream \gets getFullBitstream(incoming\_task)$
            \State $doSwap(full\_bitstream)$
            \State $setupFPGA()$
             
            \ForAll {$r \in R$}
                \If {$evicted(r) \neq r_{swap}$}
                    \State $restoreContext(evicted\_task[r])$
                    \State $launch(evicted\_task[r])$
                \EndIf
            \EndFor
        \EndIf
    \end{algorithmic}
\end{algorithm}

%The main loop of the scheduler is presented in Algorithm \ref{alg:man_loop_scheduler}, where the arrival of the next task is simulated with a timeout clock, used in the same \emph{select()} function that detects the interrupts raised by the end of a kernel in a reconfigurable region. Consequently, the \texttt{WaitForInterrupt} function returns when a new task arrives or when a RR kernel finishes.
%When a new task arrives, the module to generate it is called.
%In every iteration a task is served.
%, either just arrived/generated or retrieved from a queue, is served.

The process of serving a task consists of the following steps:
%\begin{enumerate}
%    \item 
    (1) Find an available region, i.e., a region where the last task running has already finished. \label{step:sch_available}
    (2) In case no available region was found, if preemption is disabled enqueue the task.
    %go to step \ref{step:sch_enqueue}. 
    If preemption is enabled, check if there is a region executing a task with lower priority.
    %, a task can be preempted from a region where its running task has a lower priority than the incoming task. This region is now considered available. Go to step \ref{step:sch_available}.
    In that case, stop the kernel execution in that region, save the context and state, enqueue the stopped
    task, and consider the
    region as available.
    (3) If the kernel loaded in the available region is distinct from the kernel of the incoming task, enqueue a swapping task to reconfigure the RR. \label{step:sch_swap}
    (4) Launch the new task. If it was a previously stopped task, its context is copied back to the device before
            launching.
  % \item Enqueue the incoming task and finish. \label{step:sch_enqueue}
%\end{enumerate}

%Finally, 
%The $update\_timeout()$ function loads the arrival time of the next task in the $timeout$ structure when the last incoming task has been served.

%The scheduler is modular, with separate modules for the generation of tasks, management of the queues, service of tasks and the main loop of the scheduler, therefore easy to extend. It is compatible with any number of RRs and the underlying configuration for management of partial reconfiguration is generated transparently from the kernels specified by the user, since the name of each kernel can be extracted from \texttt{CTRL\_KERNEL\_FUNCTION}.

\section{Programmer's abstractions}
\label{sec:abstractions}
Whilst dynamic partial reconfiguration is abstracted from the programmer by the FPGA shell and new Controller backend that were described in Section \ref{sec:approach}, in order to leverage preemption of kernels then some explicit support is required from the programmer. We provide high-level kernel code abstractions that allow programmers to store the state of their kernel at arbitrary points during the execution of the task, instead of waiting for a specific pre-defined consistency point. Traditional FPGA checkpointing methods, such as \textit{readback}  or \textit{scan-chain}, commonly incur either high time overhead by write the state of the whole FPGA fabric \cite{vivadodebug} or resource usage overhead by extending the existing circuit with state access support structures \cite{wheeler2001using}, respectively. Instead, by providing HLS kernel code abstractions we can provide more choice to the programmer so that they can choose when they store the state and the state storage is limited to a specific kernel.

\subsection{Kernel interface abstraction}
The generation of interfaces in technologies such as Vitis HLS is done adding pragmas that can be cumbersome and error prone to write, especially for non-experts. Moreover, as described in Section \ref{sect:architecture}, a requirement of dynamic partial reconfiguration is that HLS kernels to be deployed into a given reconfigurable region must present the same external interface to the shell. This means that they must conform to the same number of interface ports and port configurations, such as bus widths~\cite{dfx}. Consequently, abstractions to hide these low-level details to the programmer are beneficial as they simplify the overall code.

\begin{lstlisting}[language=C, basicstyle=\scriptsize, numbers=left, numbersep=8pt, xleftmargin=.25in, xrightmargin=.25in, caption={Sketch of a Median Blur kernel written with the Controller abstraction}, label={lst:median_blur}]
CTRL_KERNEL_FUNCTION(
MedianBlur, PYNQ, DEFAULT, 
KTILE_ARGS(KHitTile_int in_array,
           KHitTile_int out_array), 
INT_ARGS(int H, int W, int iters), 
FLOAT_ARGS(NO_FLOAT_ARG)) {
  ...
  int k, row, col;
  context_vars(k, row, col); @\label{code:context_vars}@
  ...
  for_save(k, 0, iters, 1) { @\label{code:for_save1}@
    for_save(row, 1, H+1, 1) { @\label{code:for_save2}@
      for_save(col, 1, W+1, 1) { @\label{code:for_save3}@
        window[0] = hit(
          in_array, row-1, H_NCOL+col-1);
        ...
        checkpoint(col); @\label{code:checkpoint1}@
      } checkpoint(row); @\label{code:checkpoint2}@
    } checkpoint(k); @\label{code:checkpoint3}@
  }   
}
\end{lstlisting}

The configuration of the interfaces is a parameter present in our Tcl configuration script that generates the shell's hardware design, as discussed in Section \ref{sect:architecture}. In the Controller model, codes forming a task are wrapped with curly brackets and preceded by a kernel signature. This kernel signature is provided with a macro-function named \texttt{CTRL\_KERNEL\_FUNCTION} which specifies the kernel parameters in a form that is processed by the Controller library to generate the proper low-level interface.
Listing~\ref{lst:median_blur} illustrates the definition of a Median Blur kernel, used in our evaluation in Section \ref{sec:experiments}, where from the the signature contains the following possible parameters:

%\texttt{CTRL\_KERNEL\_FUNCTION(K, T, S, A_{p}, A_{i}, A_{f})}: 
\begin{itemize}
\item[] \texttt{CTRL\_KERNEL\_FUNCTION(K, T, S, A\textsubscript{p}, A\textsubscript{i}, A\textsubscript{f})}: 
\begin{itemize}
     \item \texttt{K} is the name of the kernel.
     \item \texttt{T} indicates the backend type that will be targeted. Supported types are: CPU, CUDA, OpenCL, FPGA.
     \item \texttt{S} is the subtype of backend that will be targeted, e.g. \texttt{DEFAULT}.
     \item \texttt{A\textsubscript{p}} is a list of pointer non-scalar arguments defined with \texttt{KTILE\_ARGS}.
     \item \texttt{A\textsubscript{i}} is a list of integer scalar arguments defined with \texttt{INT\_ARGS}.
     \item \texttt{A\textsubscript{f}} is a list of floating point scalar arguments defined with \texttt{FLOAT\_ARGS}.
 \end{itemize}
\end{itemize}

Whilst the first three of these arguments are obvious, the last three require explanation. The \texttt{A\textsubscript{p}} argument is required because Controller provides a wrapper structure for multi-dimensional arrays named \emph{HitTile}, where any kind of non-scalar arguments are provided as \emph{HitTile} arguments. The \texttt{KTILE\_ARGS} function enables the use of \emph{HitTile} accessors within the kernel, effectively providing input and output arrays to the kernel, as discussed in \cite{rodriguez2021efficient}. The last two arguments, \texttt{INT\_ARGS} and \texttt{FLOAT\_ARGS} support passing integer and float scalar arguments, respectively and all these functions have variadic arguments to adapt the kernel interface to the number of arguments required by the programmer.

The corresponding code generated by the kernel signature for the kernel shown in Listing~\ref{lst:median_blur} is shown in Listing \ref{lst:medianblur_signature}, where three integer arguments are provided by the user and five extra dummy arguments \texttt{i\_args\_<n>} are generated. Similarly, 8 dummy floating point and 1 dummy pointer arguments are generated to fill the argument count and provide a shell compliant interface. Finally, a pointer to a \texttt{struct context} is added for context book-keeping if the task is interrupted. %and \texttt{int * return\_var} for the simulation of a return variable are%
%is generated. %This allows returning values integer values in the Controller model, that requires kernel to be \texttt{void} functions. This can be easily extended for other returning other types by extending the signature macro.

% \begin{lstlisting}[language=C, basicstyle=\scriptsize, label={lst:medianblur_signature}, caption={Code generation for the signature of the Median Blur kernel}]
% void MedianBlur(
%     wrapper_int input_array_wrapper, data_int input_array_data, 
%     wrapper_int output_array_wrapper, data_int output_array_data, 
%     wrapper_int ktile_args_0_wrapper, data_int ktile_args_0_data, 
%     int H, int W, int iters, int i_args_0, int i_args_1, int i_args_2, int i_args_3, int i_args_4, 
%     float no_float_arg, float f_args_0, float f_args_1, float f_args_2, float f_args_3, float f_args_4, float f_args_5, float f_args_6, 
%     volatile struct context * context, int * return_var);
% \end{lstlisting}
\begin{lstlisting}[language=C, basicstyle=\scriptsize, label={lst:medianblur_signature}, caption={Code generation for the signature of the Median Blur kernel}]
void MedianBlur(..., 
    int H, int W, int iters, int i_args_0, ..., int i_args_4, 
    ..., volatile struct context * context, int * return_var);
\end{lstlisting}

%The generation of interfaces in Vitis HLS happens through the use of pragmas that can be cumbersome and error prone to write. In Controller these are abstracted through the interface definition macros \texttt{DEF\_<type>\_INTERFACES}, where $type$ is one of \texttt{KTILE} (for pointer interfaces), \texttt{INT}, \texttt{FLOAT} or \texttt{RETURN}. The user provides the list of arguments for which the interface pragmas are to be generated or the keyword \texttt{NO\_<type>\_INTERFACE} if there is no argument of that type in the kernel.

%%%%
\subsection{Programmer abstractions for preemption} \label{sect:abst_preemption}
% data is easy - scalars not so! (internal anyway)

Preemption of a kernel whilst it is running requires saving its state so that it can be resumed in the future.
We propose a finer-grain approach compared to previous efforts, such as~\cite{vaishnav2019heterogeneous}, which only store the context at the end of an OpenCL workgroup. Following our programmer-aware checking approach the 
programmer has the flexibility to indicate when and what data should be checkpointed during the kernel execution.
%The kernel's external data, for instance input and output arrays, is fairly easy to handle as these are stored in explicit memory areas that can be accessed by both the FPGA and CPU, and hence conveniently accessed by both to obtain the current state of the kernel. However internal data, such as scalars, are much more difficult to handle as these are often not represented by the HLS tooling as memory, as one would expect with Von Neumann architectures, but instead might be a wire on the dataflow graph connecting one construct with another. Consequently checkpointing these is significantly more complex, with some approaches explicitly modifying the generated hardware description language to support such activities [REF]. 
%It was our objective to support preemption at the HLS level, without any changes required to the generated code. 
%Furthermore, unlike 
%Previous approaches, such as~\cite{vaishnav2019heterogeneous} only save the context at the end of an OpenCL workgroup.
%we also wanted to provide flexibility for the programmer to decide exactly where their code should be checkpointed. 
To this end we provide several checkpointing macro-functions the first of which, \texttt{context\_vars}, enables the programmer to declare which variables should be stored in the checkpoints. The \texttt{checkpoint} macro stores one or more of these variables at a given execution point, and the \texttt{for\_save} macro-function is used in-place of the normal \emph{for} loop construct, to provide support for resumption on a specific loop iteration. These macro calls are expanded at synthesis time and hence are translated to the HDL level,

%, and the programmer 
%Listing \ref{lst:median_blur} sketches the Median Blur kernel code used for our experiments in Section \ref{sec:experiments} which provides an illustration of these checkpoint-restart macros in use.
Listing \ref{lst:median_blur} provides an example of the use of these preemption macros, where  at line \ref{code:context_vars} the integer variables \emph{k}, \emph{row}, and \emph{col} are selected to be checkpointed, with lines \ref{code:for_save1}, \ref{code:for_save2}, and \ref{code:for_save3} using the \texttt{for\_save} macro to define loops and for these to be restarted as appropriate. The associated loop variables are checkpointed at lines \ref{code:checkpoint1}, \ref{code:checkpoint2}, and \ref{code:checkpoint3}. 
This kernel saves the state at each iteration so that it can be resumed without discarding previously computed
iterations. It should be noted that, in this example we have chosen to save this state at every iteration, which means that we are able to resume execution without discarding any of the previously computed iterations, although this might result in some storage overhead as described in Section \ref{sect:architecture} this is minimised by our shell using the fast on-chip BRAM memory for this checkpointing.

\begin{lstlisting}[language=C, basicstyle=\scriptsize, label={lst:struct_context}, caption={Definition of \texttt{struct context}}.]
struct context {
  int var[N];
  int init_var[N];
  int incr_var[N];
  int saved[N];
  int valid;
};
\end{lstlisting}

Context saving is undertaken transparently by storing the state of variables into the \texttt{context} structure which is then stored in BRAM. Listing~\ref{lst:struct_context} illustrates the members of this structure, where in our prototype up to \emph{N} integers can be nominated by the user to be saved (where $N$ is a compile time parameter). Whilst integers suffice for our purposes in this paper, it is trivial to extend this structure to support other data types by adding additional member(s). 

The field \texttt{saved} in Listing~\ref{lst:struct_context} stores information about whether corresponding variables have already been saved and this is read during a resume operation to determine whether they should be restored or not. The \texttt{valid} field is a safety check and used to indicate whether the kernel was interrupted during a data saving operation. This is required because preemption is asynchronous, and as such if it occurs during data saving then the stored state can be inconsistent. In such a case the resume operation will be undertaken with the previously saved values.
%and if their last value before preemption should be used. 
%If the value is 0 
%then this is the first execution of the kernel and the checkpointed 
%the variables should be set to the programmer's initial values specified in the code.
%As discussed in Section \ref{sect:architecture}, since preemption is asynchronous 

%used as a lock between code sections in which the context structure is being modified. If the field is 1 after reset, then the information in the structure is valid, otherwise it is corrupted and the last iteration will be discarded when the kernel is resumed by manipulating the values of the loop variables in the structure.

\section{Experimental study}
\label{sec:experiments}

In this study we experiment with
%The work presented in this paper is assessed through 
the scheduler described in Section \ref{sect:dpr_scheduler} to operate upon randomly generated image filter tasks. The tasks chosen for the experimentation are blur image filters applied to images pre-stored in memory, and they execute one of four possible kernels: One iteration of Gaussian Blur, or Median Blur over one, two or three iterations.
%The filters are applied over randomly generated tasks arriving at 
Tasks arrive at random times distributed over $\unif(0,T)$ minutes. With five different priorities to generate a rich mixture of task preemption, task switching and FPGA reconfiguration.
%we use and in this section we opt for five as this setting is high enough to study the effects of the scheduling on the queues.
%The tasks associated with each of our four kernels, along with their arrival time, and input and output images are randomly generated before the scheduler starts. 
%The tasks associated with each of our four kernels, along with their arrival time and input and output images, are randomly generated before the scheduler starts. 
The tasks, their arrival time, and the image on which it should be applied, are randomly generated before the scheduler starts.

\subsection{Experimentation environment}
The experiments were conducted on a Xilinx PYNQ-Z2 FPGA featuring a ZYNQ XC7Z020-1CLG400C of the Zynq-7020 family, an ARM Cortex-A9 dual core at 650 MHz CPU and 512 MB DDR3. The FPGA fabric is running at a clock frequency of 100MHz and HLS kernels were compiled using Xilinx Vitis HLS version 2020.2 with the hardware design and corresponding bitstreams generated with Xilinx Vivado v2020.2. Controller was compiled with GCC 7.3.0 and compilation scripts were generated with CMake 3.25.1.

Thirty tasks are generated for each experiment, where each task has a randomly chosen priority value from 0 to 4 with the same probability, with 0 being the highest priority. 
The kernel to be executed by the task is also randomly chosen with the same probability among the four possible kernels and we consider three different scenarios of service load by changing  the maximum possible arrival time $T$ of the tasks. This changes the average arrival rate of the tasks, where in the \emph{Busy} scenario we use $T=0.1$ minutes, in the \emph{Medium} scenario $T=0.5$ minutes and in the \emph{Idle} scenario $T=0.8$ minutes.
%We enabled priorities both with and without preemption of tasks. We considered three different rate of arrivals $T$: busy (0.1), medium (0.5) and idle (0.8).
All the tasks details (arrival time, kernel to execute and priority) are  randomly chosen with a Tausworthe random generator initialised with a given seed for experiment reproducibility. We present results for 10 seeds of a Tausworthe random sequence starting at an arbitrarily chosen seed value: 28871727. The main observations can be extrapolated for other random sequences.

We have also designed experiments using different image sizes to generate different task loads distributions, and in this section we explore image sizes that are $200\times200$, $300\times300$, $400\times400$, $500\times500$ and $600\times600$. To study the scheduling behaviour of the tasks executing sequentially, compared to running in parallel, both one and two reconfigurable region scenarios are considered. We also study the effect of using scheduling policies with and without preemption, where preemption is activated when an arriving task has higher priority than one of the tasks currently in execution.

%for the results for service time, and this is defined as the amount of time taken for a task to start execution on a RR after its arrival. This is useful as it illustrates an important scenario in which preemption is beneficial due to the priorities of the incoming tasks. Although not shown here the scheduler was tested with a range of seeds to ensure generality. 
%We worked with image sizes $200\times200$, $300\times300$, $400\times400$, $500\times500$ and $600\times600$. In order to study the sequential vs. the parallel behaviour both one and two RRs were considered. \textcolor{blue}{Finally, each experiment was executed ten times to account for variability and the results presented are average times with standard deviation.}

\subsection{Resource usage}Our first study involved exploring the resource usage of our two kernels, the Median Blur and Gaussian Blur with and without support for preemption enabled. The purpose of this was to understand the FPGA resource overhead resulting from enabling preemption, and Table \ref{table:resource_usage} presents these figures as reported by the HLS tooling. It can be seen that, as expected, with the Median Blur kernel when enabling preemption this increases the resource usage. However, surprisingly with the more complicated Gaussian Blur kernel when enabling preemption this reduces the overall usage rather than increases it. By analysing the logs and resource usage reports that are generated by the tooling we found that the HLS tool follows a different synthesis path for the Gaussian Blur kernel when preemption is enabled, generating different HDL target code that is fairly significantly different from when preemption is disabled. Nevertheless, it can be seen from Table \ref{table:resource_usage} that the resource overhead of enabling preemption is minimal, especially for the LUTs. Incidentally, the reason we are not reporting BRAM usage in Table \ref{table:resource_usage} is because, as described in Section \ref{sect:architecture} this sits external to the HLS kernel in the shell and can be sized as required to fit within the constrains of the available memory on the chip.

\begin{table*}[htb]
\centering
\caption{Single HLS kernel FPGA resource usage for Median Blur and Gaussian Blur when enabling or disabling preemption.}
\label{table:resource_usage}
\begin{tabular}{ccc}
\toprule
Description & LUT resource usage \% & DSP resource usage \%\\
\midrule
Median Blur (no preemption) & 0 & 1 \\
Median Blur (preemption) & 1 & 14 \\
Gaussian Blur (no preemption) & 21 & 12 \\
Gaussian Blur (preemption) & 17 & 3 \\
\bottomrule
\end{tabular}
\end{table*}

\subsection{Results}
\begin{figure}
    \centering
    \begin{subfigure}{0.48\textwidth}
    \includegraphics[width=\textwidth]{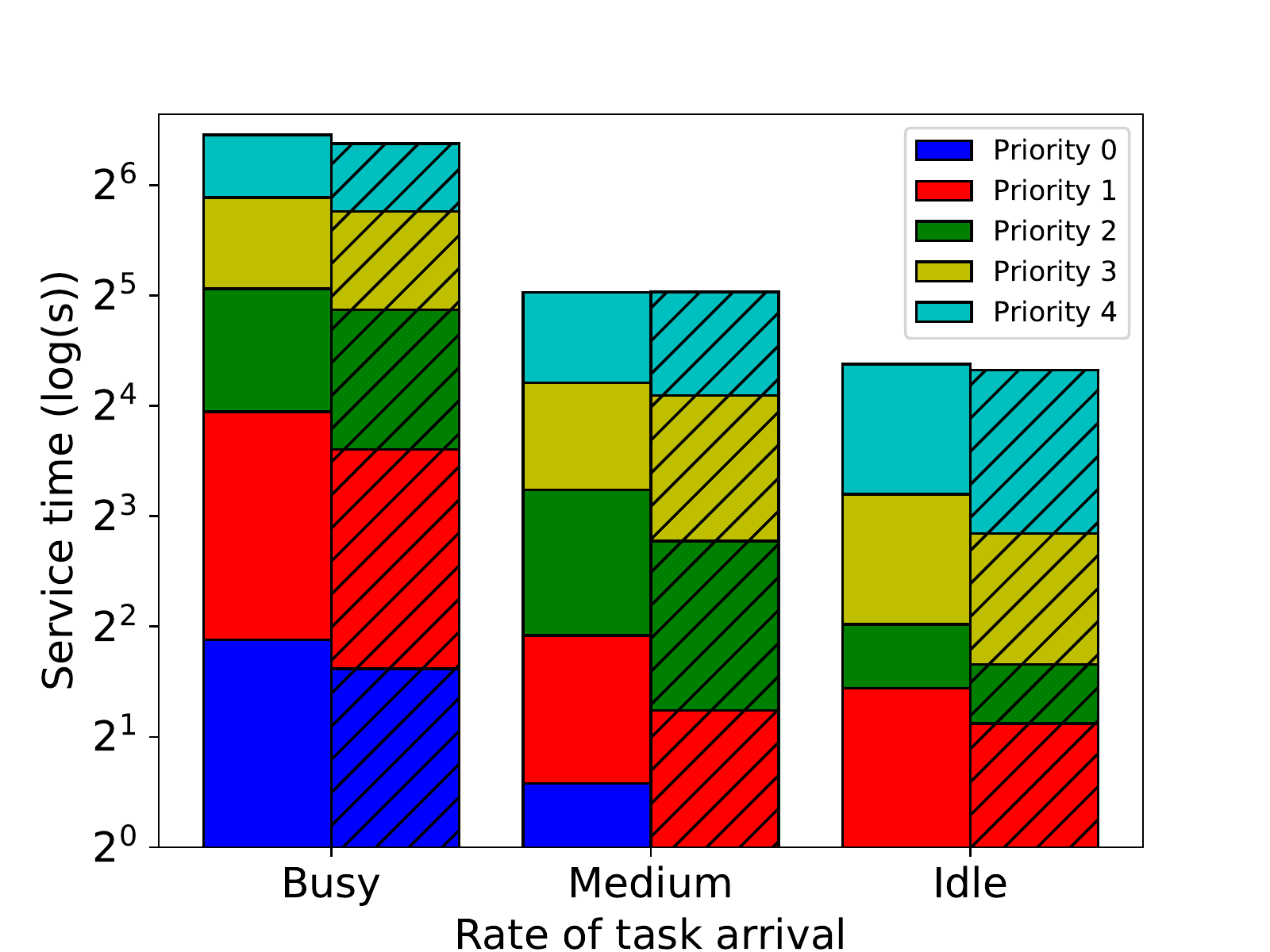}
    \label{fig:service_time_1rr}
    \end{subfigure}
    \begin{subfigure}{0.48\textwidth}
    \includegraphics[width=\textwidth]{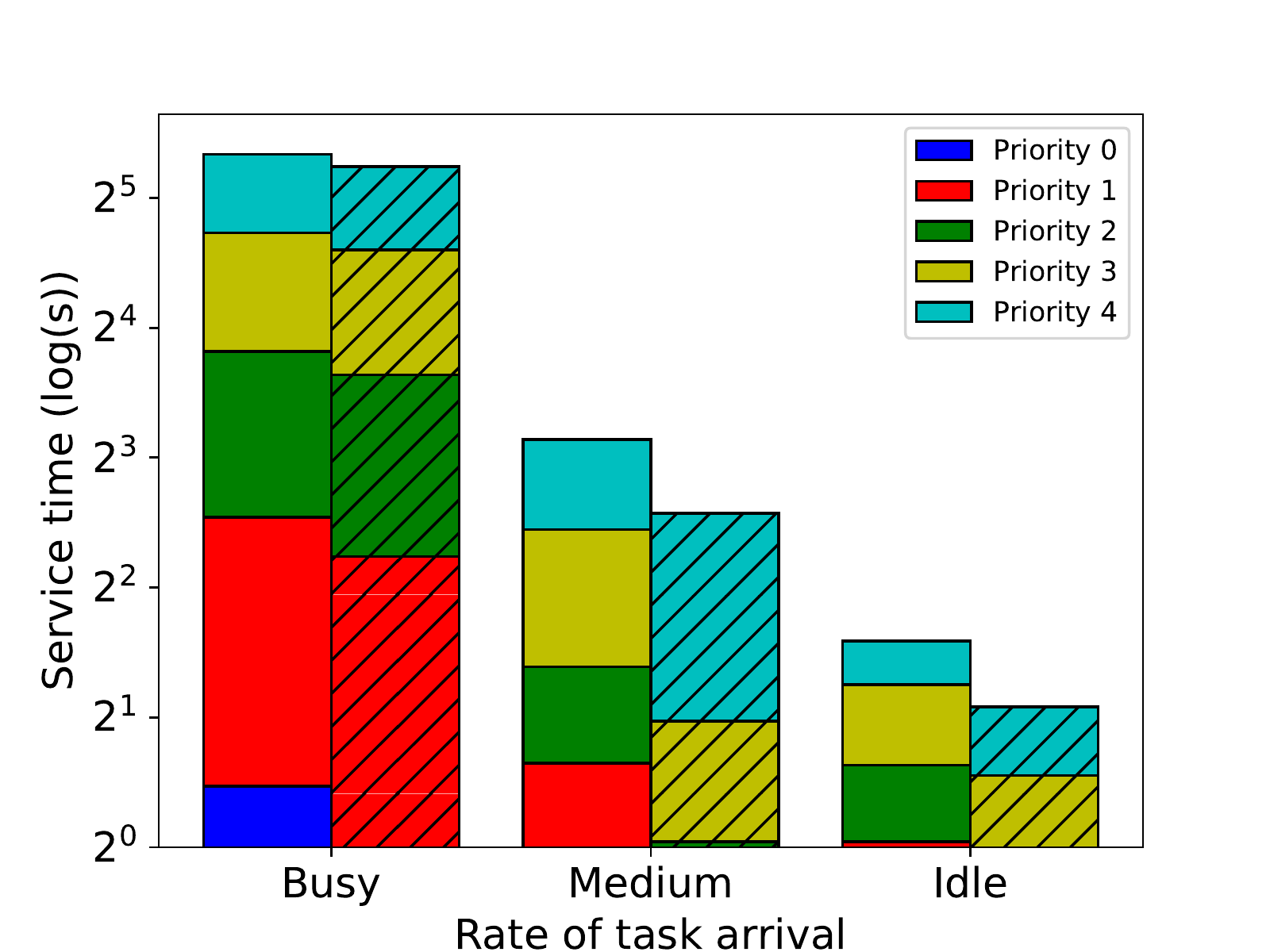}
    \label{fig:service_time_2rr}
    \end{subfigure}
    \caption{Service times for 30 tasks at size $600 \times 600$ for seed 1120249751, with one reconfigurable region in the diagram on the left, and two regions on the right. The bars in each plot represent non-preemptive execution on the left and preemptive execution on the right.}
    \label{fig:service_times}
\end{figure}

In order to show the effectiveness of our approach we are presenting results for the following metrics: (i) service time, defined as the time it takes for a task to be served since it is generated until it starts execution on the FPGA and (ii) throughput, defined as the number of tasks executed per second. %Note that (i) only measures the time for the first service and does not take into account the time for further services after preemption, since this metric intends to measure the effectiveness of the scheduler to serve urgent tasks. 
We also compare the use of partial reconfiguration with the more conventional full reconfiguration approach.
Figure \ref{fig:service_times} reports the service time for tasks in every priority queue both with and without preemption for 30 tasks at size $600\times600$ accumulated by priority. We chose this number of tasks and image size as it provides a large enough workload and a sufficient number of tasks to study the behaviour of the scheduler. The results are presented both for one and two reconfigurable regions and, as can be seen, service times are longer for the busy rate of arrival than for medium and idle, as tasks have to wait a longer time until a region becomes available than when they arrive later, providing the opportunity for kernels to finish. 

If the priority of an incoming task is higher than one of the tasks running, then its service time will be virtually zero. We can observe this by comparing the right bars of Figure \ref{fig:service_times} with the corresponding left bars. For this representative case, on average, preemption reduces service time substantially and this will be the case in general when incoming tasks present a higher priority than running tasks. These results demonstrate that our scheduler effectively reduces the total service time of tasks, thus increasing the flexibility, as preemption enables swapping in and out tasks upon a specific condition which, in this case, is the priority. The reduction in service time is heavily dependent on the structure of priorities of the generated tasks, both in terms of the number of tasks enqueued and the number of reconfigurations required by the incoming kernels which are not already loaded in the fabric. 
Note that a task will have to wait until previous tasks of higher or the same priority have completed.
%This can create the effect that even highest priority tasks may starve, as other highest priority tasks might have arrived before. 
Additionally, as shown in Figure \ref{fig:service_times}, the total service time decreases with the number of reconfigurable regions, as more opportunities are created for kernels of lower priorities to execute.
%that can be evicted.
%and available RRs are created.

% \begin{figure}[thb]
% \begin{subfigure}{0.5\textwidth}
%     %\centering
%     \includegraphics[width=\textwidth]{plots/service_times-no-preemption-1rr-600-15.pdf}
%     \caption{1 RR without preemption.}
%     \label{fig:my_label}
% \end{subfigure}
% \begin{subfigure}{0.5\textwidth}
%     %\centering
%     \includegraphics[width=\textwidth]{plots/service_times-preemption-1rr-600-15.pdf}
%     \caption{1 RR with preemption.}
%     \label{fig:my_label}
% \end{subfigure}
% % \caption{Service times for 30 tasks launched on 1 RR.}
% % \end{figure}

% % \begin{figure}
% \begin{subfigure}{0.5\textwidth}
%     %\centering
%     \includegraphics[width=\textwidth]{plots/service_times-no-preemption-2rr-600-15.pdf}
%     \caption{2 RR without preemption.}
%     \label{fig:my_label}
% \end{subfigure}
% \begin{subfigure}{0.5\textwidth}
%     %\centering
%     \includegraphics[width=\textwidth]{plots/service_times-preemption-2rr-600-15.pdf}
%     \caption{2 RR with preemption.}
%     \label{fig:my_label}
% \end{subfigure}
% \caption{Service times for 30 tasks at size $600\times600$.}
% \label{fig:service_times}
% \end{figure}

\begin{figure*}
    \centering
    \begin{subfigure}{0.48\textwidth}
    \includegraphics[width=\textwidth]{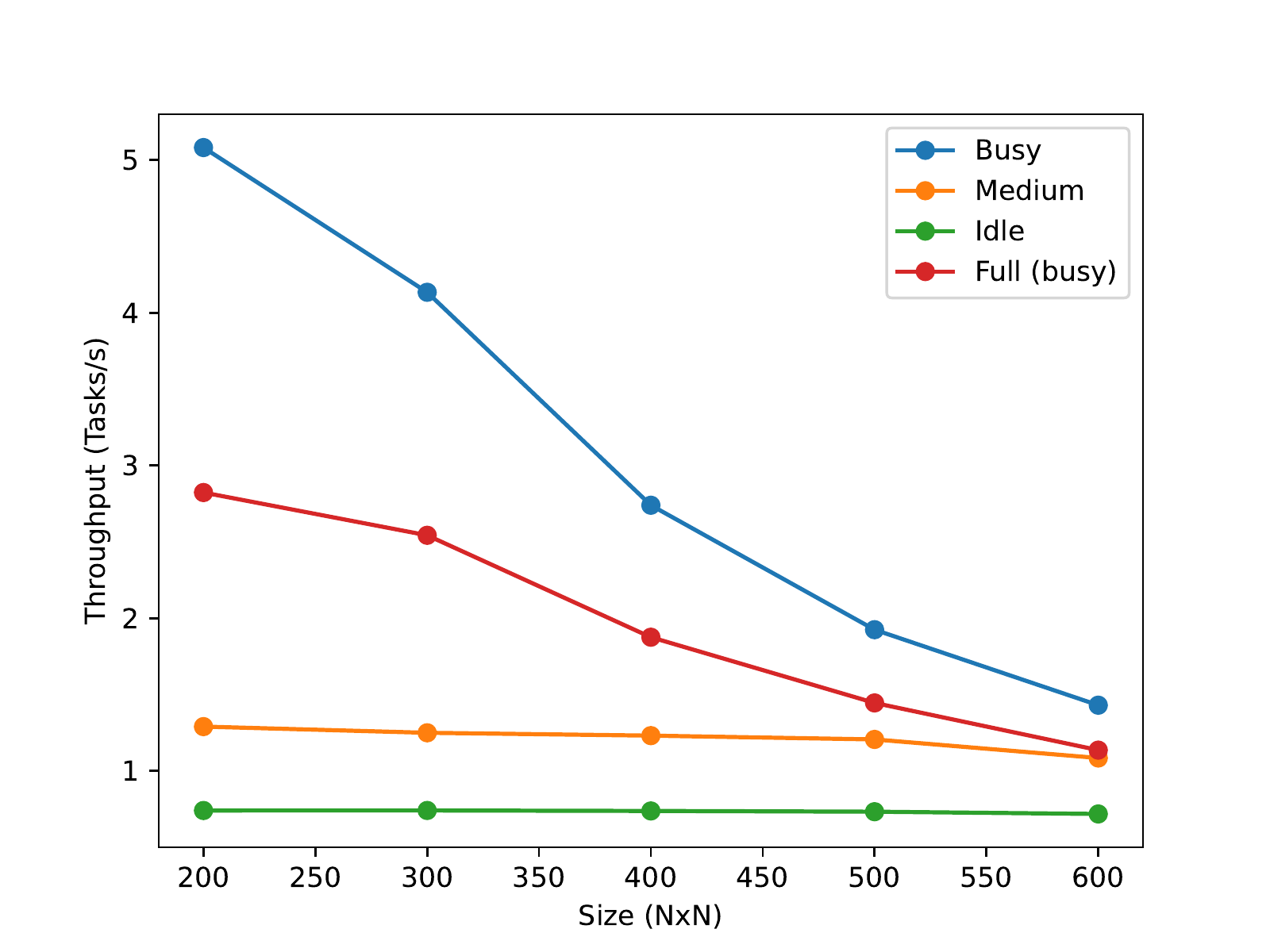}
    \label{fig:throughput_28871727}
    \end{subfigure}
    \begin{subfigure}{0.48\textwidth}
    \includegraphics[width=\textwidth]{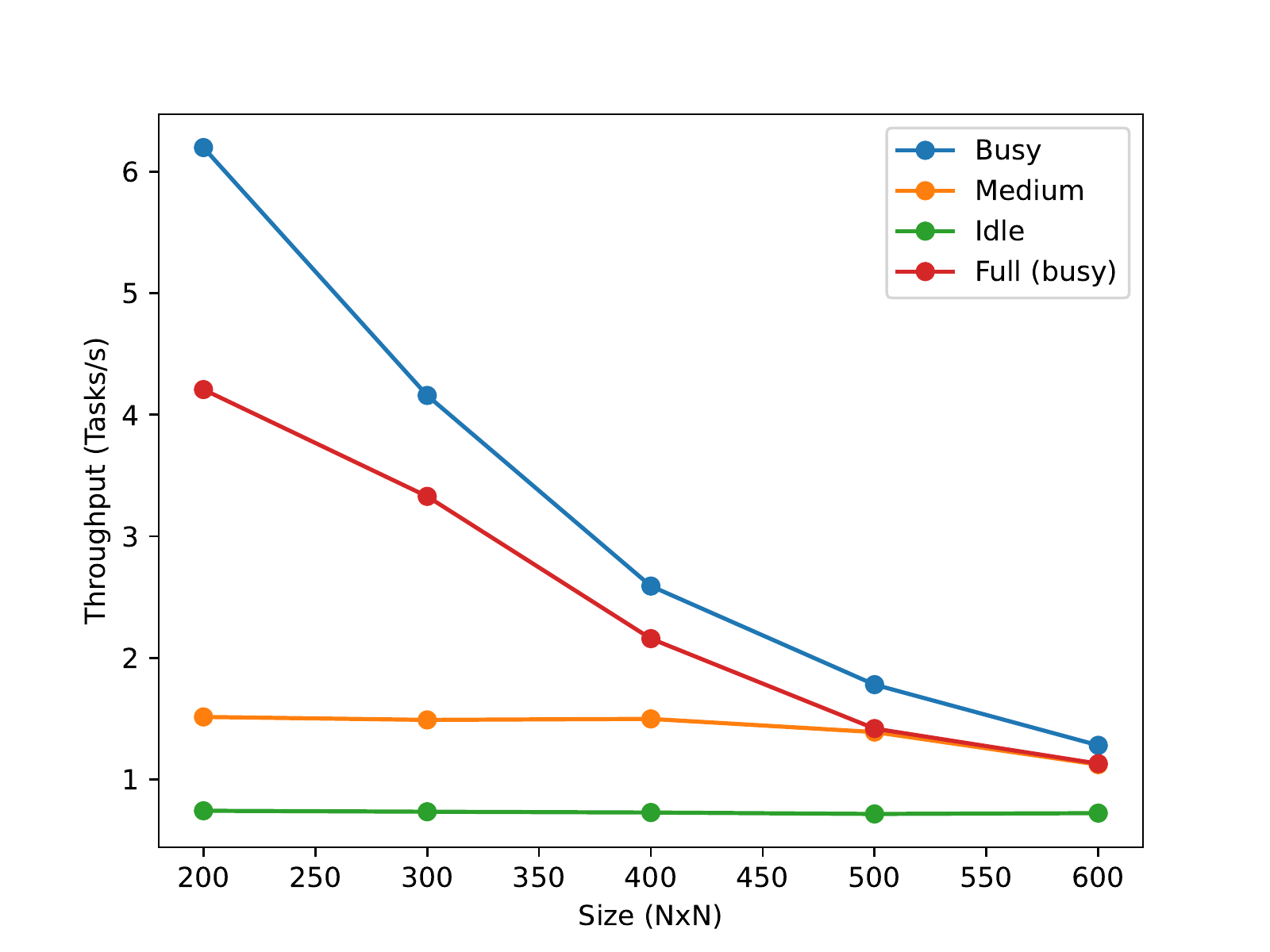}
    \label{fig:throughput_1368297677}
    \end{subfigure}
    \begin{subfigure}{0.48\textwidth}
    \includegraphics[width=\textwidth]{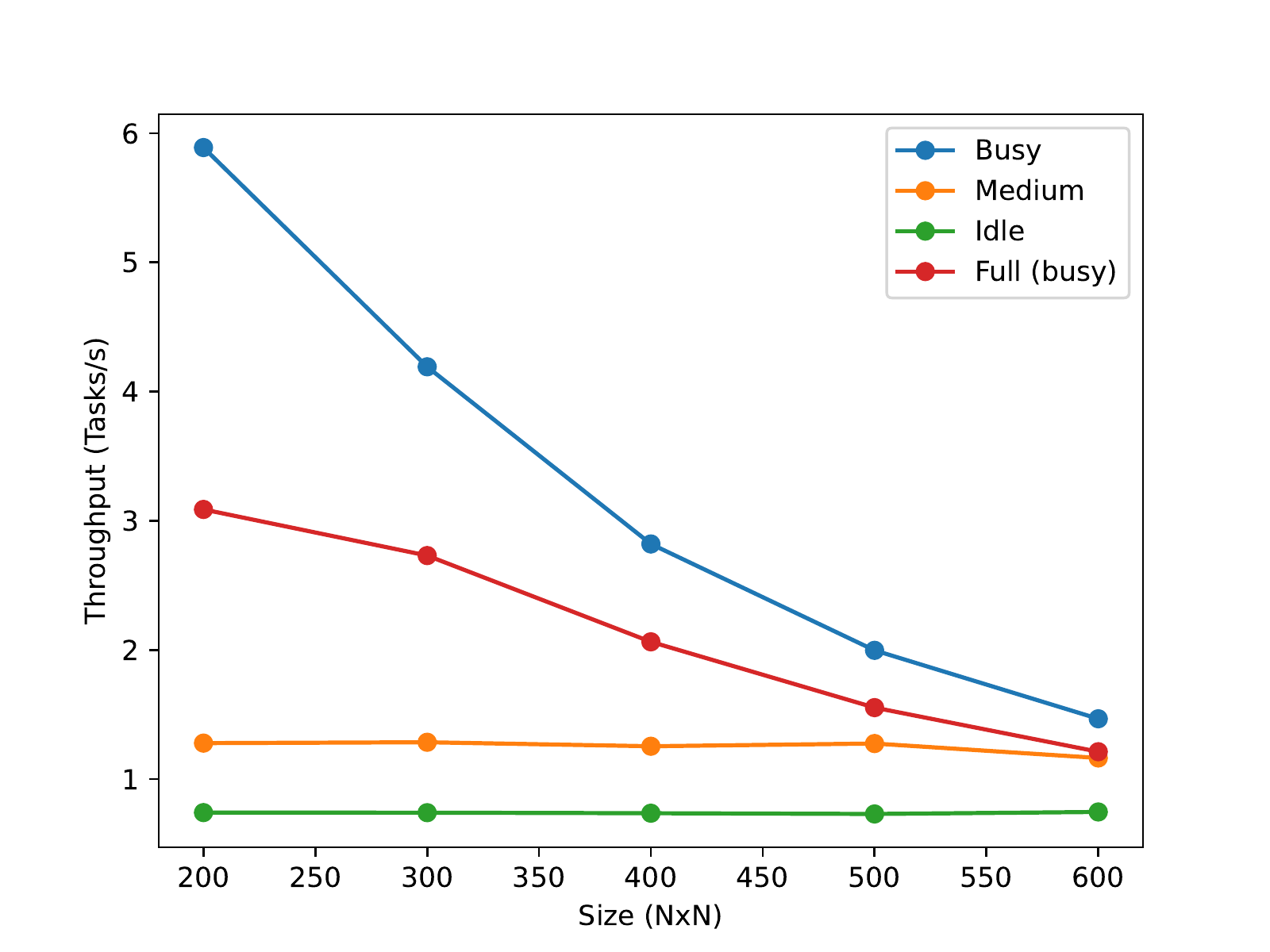}
    \label{fig:throughput_28871727}
    \end{subfigure}
    \begin{subfigure}{0.48\textwidth}
    \includegraphics[width=\textwidth]{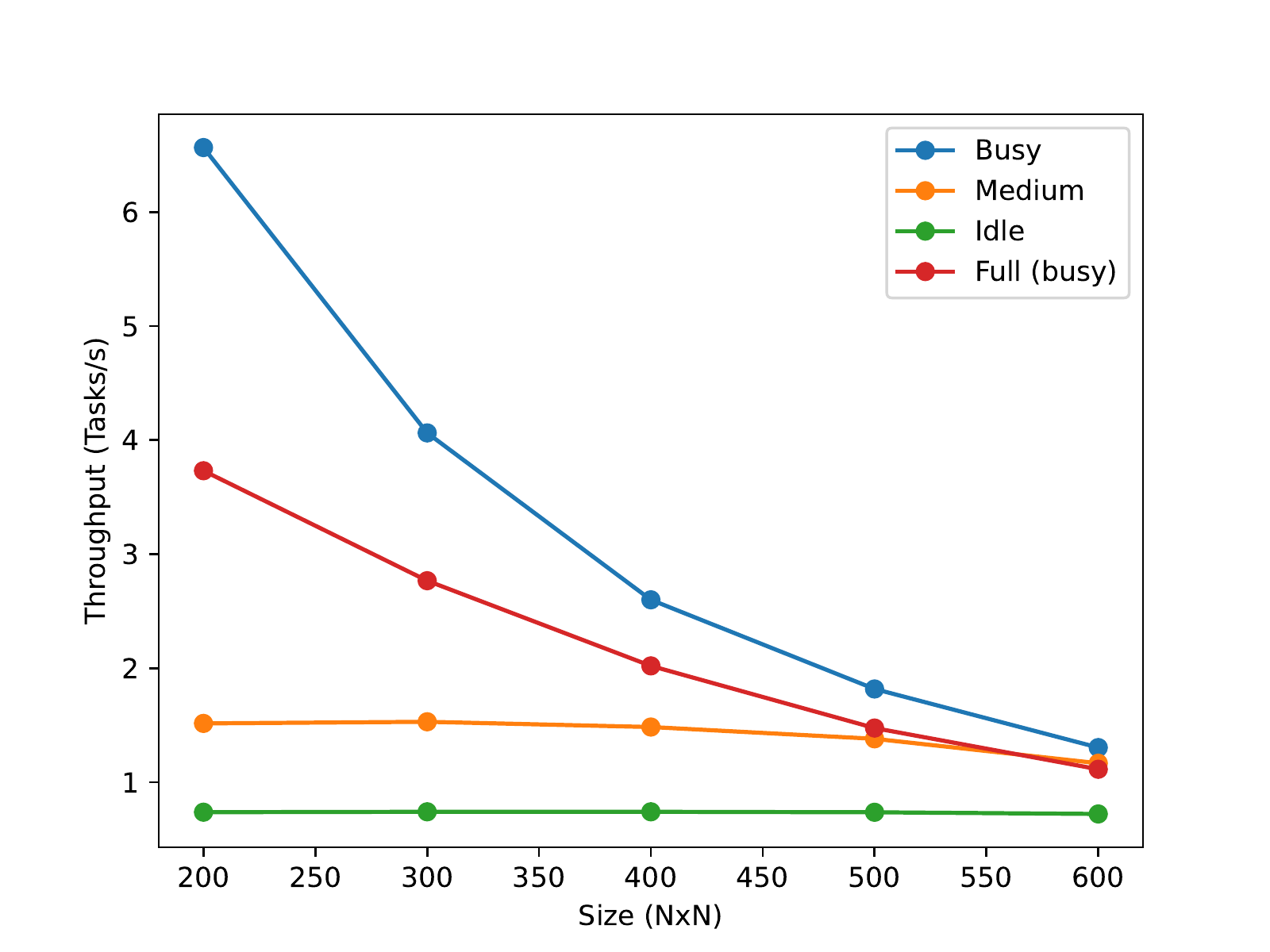}
    \label{fig:throughput_28871727}
    \end{subfigure}
    \caption{Throughput for 30 tasks scheduled with preemption (top) and without preemption (bottom). Seed 28871727 (left) and 1368297677 (right).}
    \label{fig:throughput}
\end{figure*}

Figure \ref{fig:throughput} illustrates the throughput of the scheduler for 30 tasks with preemption and dynamic partial reconfiguration over two reconfigurable regions for two of the randomly chosen seeds. The throughput of the preemptive scheduling with full reconfiguration in the busy scenario is the highest achieved with this policy and is provided for comparison in Figure \ref{fig:throughput}. As expected, the throughput increases with the rate of arrival of tasks, and as a general pattern the lower the dimensions of the images then the higher the throughput, as kernels complete their execution faster. It is also observable that the overheads incurred by preemption lead to a slightly lower overall throughput. These are most noticeable for a high rate of arrival of tasks, where throughput losses are 13.71\% and 5.58\% for the case with seed 28871727 and 1368297677, respectively, at size 200 and busy arrival rate.
%are observed.
For the rest of cases the loss ranges between 1.38\% and 3.60\% for the seed 28871727 and -2.37\% and 3.40\% for the seed 1368297677. Most of this overhead is explained by the time taken by the extra partial reconfigurations required by preemption. The red lines in the graphs of Figure \ref{fig:throughput} report the throughput when full reconfiguration is instead used. In this situation it can be seen that full reconfiguration performs much worse than dynamic partial reconfiguration, even in the most favourable of case.

\begin{figure}
    \centering
    \begin{subfigure}{0.48\textwidth}
    \includegraphics[width=\textwidth]{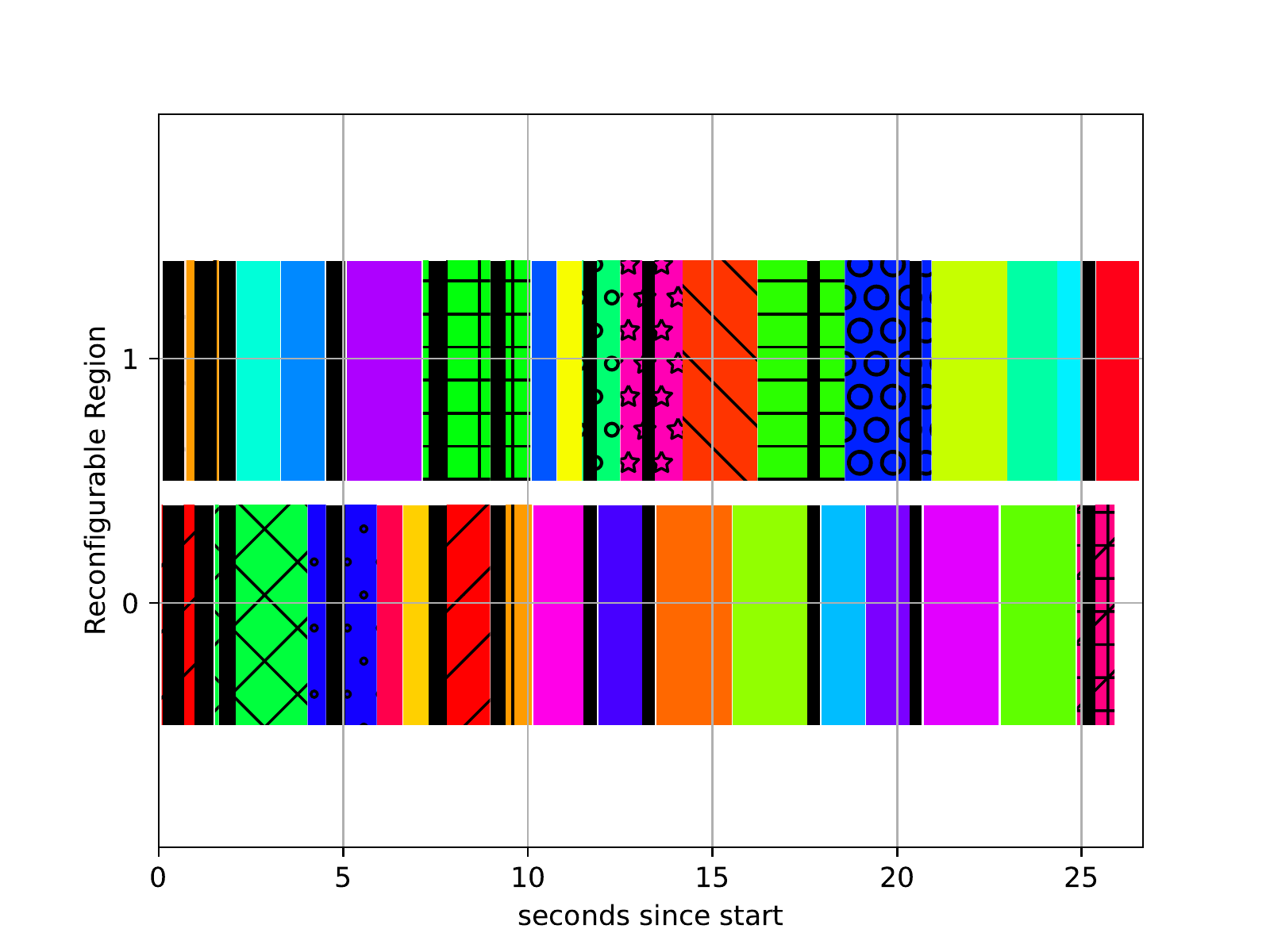}
    \label{fig:gantt_full}
    \end{subfigure}
    \begin{subfigure}{0.48\textwidth}
    \includegraphics[width=\textwidth]{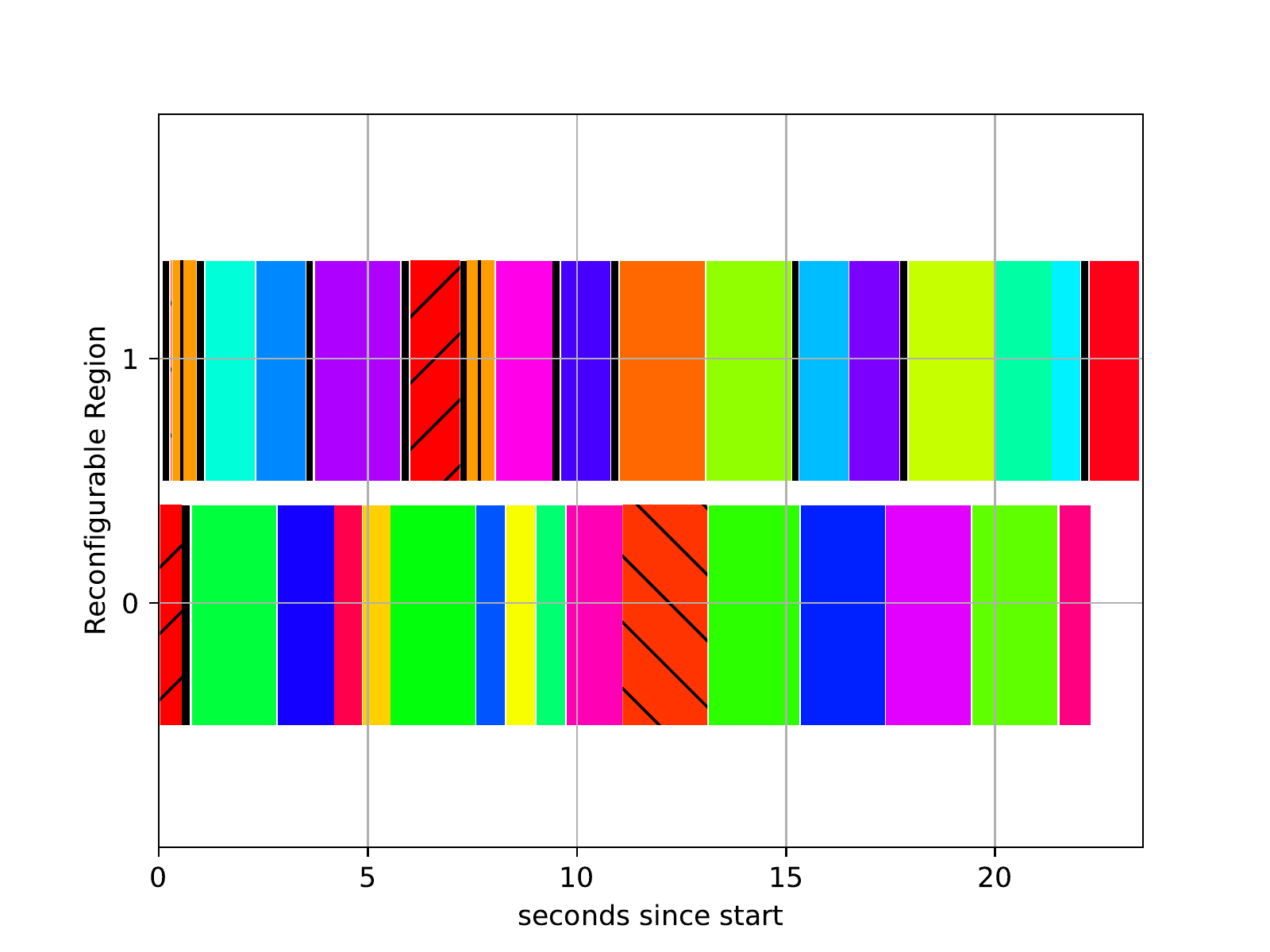}
    \label{fig:gantt_partial}
    \end{subfigure}
    \caption{Scheduling of 30 tasks at size $600 \times 600$ over two reconfigurable regions for seed 1368297677. Full reconfiguration is on the left and partial reconfiguration on the right. Execution tasks shown in colour and swap operations in black. Preempted tasks present a hatched pattern.}
    \label{fig:gantt}
\end{figure}

Figure \ref{fig:service_times} reports results from running an experiment with 30 tasks working with images of $600 \times 600$ pixels on two reconfigurable regions for the seed 1368297677. The figure shows the results for full reconfiguration (left), and dynamic partial reconfiguration (right). Execution tasks are shown in colour and swap operations in black. Preempted tasks are presented with a hatched pattern.

\begin{table}[htb]
\centering
\caption{Average service time for maximum priority with standard deviation for 30 tasks scheduled on 1 RRs at size $600\times600$. $\overline{T_m}$ is the average service time, where T can be B (busy), M (medium) or I (idle) and m can be P (preemptive) or NP (non-preemptive).}
\label{table:service_time_p0_1}
\begin{tabular}{ccccccc}
\toprule
seed & $\overline{B_{np}}$ & $\overline{M_{np}}$ & $\overline{I_{np}}$ & $\overline{B_p}$ & $\overline{M_p}$ & $\overline{I_p}$ \\
\midrule
28871727 & 4.61 & 0.63 & 0.58 & 4.12 & 0.25 & 0.14 \\
1368297677 & 5.16 & 3.18 & 1.44 & 4.12 & 1.77 & 0.41 \\
3968565823 & 5.82 & 2.73 & 2.03 & 3.82 & 1.24 & 0.90 \\
1120249751 & 3.69 & 1.50 & 0.97 & 3.07 & 0.76 & 0.62 \\
3706141637 & 5.83 & 2.74 & 2.03 & 3.82 & 1.25 & 0.90 \\
1838770479 & 3.98 & 2.36 & 1.32 & 3.36 & 0.92 & 0.28 \\
980516246 & 4.60 & 0.65 & 0.98 & 3.20 & 0.29 & 0.30 \\
407297508 & 3.40 & 1.44 & 0.81 & 1.38 & 0.34 & 0.31 \\
3820789643 & 5.79 & 2.72 & 2.05 & 3.82 & 1.23 & 0.90 \\
1227911765 & 2.88 & 0.80 & 0.32 & 2.86 & 0.16 & 0.06 \\
\bottomrule
\end{tabular}
\end{table}

In the case of full reconfiguration, the kernel-code swap  operation is executed in both reconfigurable regions concurrently, and therefore the whole FPGA fabric remains idle during reconfiguration. Figure \ref{fig:service_times}-left shows that all the running tasks have to be preempted before full reconfiguration and restored afterwards. This process, in addition to the necessary book-keeping, incurs large overheads. It can also be seen that the reconfiguration time of full reconfiguration is much larger than that of dynamic partial reconfiguration, consequentially penalising the overall execution time. As a result of these performance penalties, the execution time with our approach is always shorter.

\begin{table}[htb]
\centering
\caption{Average service time for maximum priority with standard deviation for 30 tasks scheduled on 2 RRs at size $600\times600$. $\overline{T_m}$ is the average service time, where T can be B (busy), M (medium) or I (idle) and m can be P (preemptive) or NP (non-preemptive).}
\label{table:service_time_p0_2}
\begin{tabular}{ccccccc}
\toprule
seed & $\overline{B_{np}}$ & $\overline{M_{np}}$ & $\overline{I_{np}}$ & $\overline{B_p}$ & $\overline{M_p}$ & $\overline{I_p}$ \\
\midrule
28871727 & 2.76 & 0.32 & 0.26 & 1.72 & 0.08 & 0.10 \\
1368297677 & 3.05 & 1.32 & 0.40 & 1.47 & 0.38 & 0.09 \\
3968565823 & 2.72 & 1.05 & 0.79 & 1.47 & 0.26 & 0.09 \\
1120249751 & 1.39 & 0.84 & 0.64 & 0.78 & 0.22 & 0.21 \\
3706141637 & 2.71 & 0.95 & 0.78 & 1.47 & 0.25 & 0.10 \\
1838770479 & 2.21 & 1.02 & 0.50 & 1.19 & 0.03 & 0.09 \\
980516246 & 2.39 & 0.94 & 0.85 & 1.18 & 0.10 & 0.14 \\
407297508 & 1.92 & 0.41 & 0.11 & 0.43 & 0.08 & 0.04 \\
3820789643 & 2.67 & 1.00 & 0.79 & 1.46 & 0.27 & 0.10 \\
1227911765 & 1.47 & 0.27 & 0.11 & 0.98 & 0.14 & 0.07 \\
\bottomrule
\end{tabular}
\end{table}

In the case of dynamic partial reconfiguration, whilst a reconfigurable region is executing either a kernel task or a kernel-code swap operation (indicated by black bands in the plot), the other region may be executing another kernel task with the same or different kernel code. 
In this scenario, more than one kernel can coexist on the FPGA and therefore there are no times when the reconfigurable regions are idle. Consequently, each reconfigurable region is always executing tasks or kernel-code swap operations, and by eliminating idle times, the execution time of the application is minimised. It can be seen that this technique makes better use of the FPGA resources than full reconfiguration.

\begin{table}[htb]
\centering
\caption{Average service time for minimum priority with standard deviation for 30 tasks scheduled on 1 RRs at size $600\times600$. $\overline{T_m}$ is the average service time, where T can be B (busy), M (medium) or I (idle) and m can be P (preemptive) or NP (non-preemptive).}
\label{table:service_time_p4_1}
\begin{tabular}{ccccccc}
\toprule
seed & $\overline{B_{np}}$ & $\overline{M_{np}}$ & $\overline{I_{np}}$ & $\overline{B_p}$ & $\overline{M_p}$ & $\overline{I_p}$ \\
\midrule
28871727 & 35.02 & 23.22 & 8.13 & 35.21 & 26.63 & 8.99 \\
1368297677 & 39.28 & 28.00 & 7.29 & 40.15 & 28.12 & 7.90 \\
3968565823 & 27.06 & 22.75 & 17.10 & 27.08 & 23.58 & 13.70 \\
1120249751 & 28.67 & 14.18 & 11.64 & 28.86 & 15.66 & 12.83 \\
3706141637 & 27.09 & 22.69 & 17.10 & 27.10 & 23.80 & 13.70 \\
1838770479 & 39.45 & 31.77 & 4.09 & 39.45 & 32.12 & 4.55 \\
980516246 & 30.45 & 19.01 & 4.77 & 30.33 & 19.10 & 7.22 \\
407297508 & 30.11 & 23.61 & 4.47 & 29.85 & 24.97 & 4.84 \\
3820789643 & 27.02 & 22.54 & 17.14 & 27.31 & 23.53 & 13.75 \\
1227911765 & 32.86 & 18.79 & 3.21 & 32.77 & 20.59 & 2.80 \\
\bottomrule
\end{tabular}
\end{table}

In the case of dynamic partial reconfiguration, the time required by the kernel-code swap 
 operations is proportional to the size of the region. However, for the full reconfiguration alternative the time required is proportional to  the size of the whole FPGA. Therefore it can be seen that full reconfiguration is not recommended in this sort of workload since the reconfiguration time can increase with each new generation of FPGAs.

 \begin{table}[htb]
\centering
\caption{Average service time for minimum priority with standard deviation for 30 tasks scheduled on 2 RRs at size $600\times600$. $\overline{T_m}$ is the average service time, where T can be B (busy), M (medium) or I (idle) and m can be P (preemptive) or NP (non-preemptive).}
\label{table:service_time_p4_2}
\begin{tabular}{ccccccc}
\toprule
seed & $\overline{B_{np}}$ & $\overline{M_{np}}$ & $\overline{I_{np}}$ & $\overline{B_p}$ & $\overline{M_p}$ & $\overline{I_p}$ \\
\midrule
28871727 & 17.16 & 4.96 & 0.52 & 17.66 & 5.81 & 0.35 \\
1368297677 & 19.46 & 6.08 & 0.16 & 19.50 & 6.54 & 0.18 \\
3968565823 & 13.45 & 7.32 & 1.93 & 13.25 & 7.01 & 1.88 \\
1120249751 & 13.86 & 3.37 & 0.62 & 13.56 & 3.98 & 0.65 \\
3706141637 & 13.30 & 7.40 & 1.93 & 13.35 & 5.07 & 1.89 \\
1838770479 & 19.99 & 1.44 & 0.63 & 20.08 & 1.64 & 0.94 \\
980516246 & 14.80 & 0.66 & 0.48 & 14.75 & 0.70 & 0.47 \\
407297508 & 12.54 & 2.32 & 1.19 & 12.70 & 2.61 & 1.19 \\
3820789643 & 13.27 & 7.22 & 1.93 & 13.28 & 5.02 & 1.89 \\
1227911765 & 16.21 & 1.56 & 0.34 & 15.87 & 1.80 & 0.34 \\
\bottomrule
\end{tabular}
\end{table}

 \begin{table*}[htb]
\centering
\caption{Average throughput with standard deviation for 30 tasks scheduled on 2 RRs at size $600\times600$. $\overline{T}$ is the average throughput, where T can be B (busy), M (medim) or I (idle).}
\label{table:throughput}
\begin{tabular}{ccccccc}
\toprule
size & $\overline{B_{np}}$ & $\overline{M_{np}}$ & $\overline{I_{np}}$ & $\overline{B_p}$ & $\overline{M_p}$ & $\overline{I_p}$ \\
\midrule
200 & 6.87±1.16 & 1.42±0.12 & 0.79±0.04 & 6.27±1.23 & 1.42±0.12 & 0.79±0.04 \\
300 & 4.53±0.50 & 1.42±0.13 & 0.79±0.04 & 4.26±0.41 & 1.40±0.13 & 0.78±0.04 \\
400 & 2.94±0.25 & 1.41±0.13 & 0.79±0.05 & 2.86±0.27 & 1.39±0.14 & 0.78±0.05 \\
500 & 2.02±0.13 & 1.36±0.12 & 0.78±0.04 & 1.97±0.14 & 1.34±0.15 & 0.77±0.05 \\
600 & 1.46±0.09 & 1.27±0.15 & 0.77±0.05 & 1.44±0.09 & 1.23±0.18 & 0.77±0.05 \\
\bottomrule
\end{tabular}
\end{table*}

\begin{table*}[htb]
\centering
\caption{Average overhead with standard deviation for 30 tasks scheduled on 2 RRs at size $600\times600$. Preemptive vs non-preemptive with partial reconfiguration (left), preemptive with full reconfiguration vs preemptive with partial reconfiguration (right). $\overline{T}$ is the average overhead for preemption with partial reconfiguration and $\overline{F_T}$ is the overhead of preemption with full reconfiguration, where T can be B (busy), M (medium), I (idle).}
\label{table:overhead}
\begin{tabular}{cccc|ccc}
\toprule
size & $\overline{B}$ & $\overline{M}$ & $\overline{I}$ & $\overline{F_B}$ & $\overline{F_M}$ & $\overline{F_I}$ \\
\midrule
200 & 0.10±0.05 & 0.00±0.01 & -0.00±0.01 & 0.25±0.30 & 0.27±0.23 & 0.28±0.22 \\
300 & 0.06±0.07 & 0.02±0.01 & 0.01±0.01 & 0.28±0.27 & 0.28±0.23 & 0.28±0.22 \\
400 & 0.03±0.02 & 0.02±0.02 & 0.01±0.01 & 0.27±0.24 & 0.28±0.22 & 0.28±0.21 \\
500 & 0.03±0.02 & 0.02±0.03 & 0.01±0.02 & 0.26±0.22 & 0.27±0.21 & 0.27±0.21 \\
600 & 0.01±0.01 & 0.04±0.03 & 0.01±0.01 & 0.24±0.21 & 0.26±0.21 & 0.27±0.20 \\
\bottomrule
\end{tabular}
\end{table*}

Tables \ref{table:service_time_p0_1} and \ref{table:service_time_p0_2} report the average service time for maximum priority tasks for busy, medium, and idle scenarios using preemptive and non-preemptive scheduling options in partial reconfiguration cases.  Similarly, \ref{table:service_time_p4_1} and \ref{table:service_time_p4_2} present the same results for the minimum priority. 
Different seeds, randomly obtained from a Tausworthe generator, were used to select the arrival times and task priority. The number of tasks executed is always 30, scheduled on both one and two reconfigurable regions respectively. These tables demonstrate that, that as we displace from a busy scenario to a medium or idle scenario, the service time is reduced. Consequently, less stressed scenarios obtain better results even for lower priority tasks, as there is a higher chance that a reconfigurable region is available upon task arrival.

For any of the scenarios (busy, medium or busy ) all the preemptive cases obtain better results than the non-preemptive ones. Therefore, highest priority tasks are served earlier if the preemptive policy is exploited. Adding a second reconfigurable region significantly improves service time  of maximum priority tasks since more than one task can be executed at the same time on the FPGA. In conclusion, it is highly beneficial to increase the number of reconfigurable regions to as many as can be supported by the hardware resources.

Table \ref{table:overhead} reports the average overhead of our preemption method with dynamic partial reconfiguration (left) and full reconfiguration (right). The overheads for dynamic partial reconfiguration were calculated as quotients of the average throughput of the experiments with 10 random seeds, which are presented in Table \ref{table:throughput}. In the case of full reconfiguration, these overheads were calculated as quotients with the busy rate of arrival with preemption as baseline. We tested the scheduler in scenarios involving tasks working in different image sizes to study the overhead of the preemption technique with different task loads. The overhead for dynamic partial reconfiguration is larger for busier scenarios, as the rapid arrival of tasks triggers more kernel preemptions. Similarly, the overhead is also greater for smaller 
image sizes since the execution time is shorter and the throughput is higher, leading to more 
reconfigurations. Note that the worst overhead obtained is 10 $\pm$ 5 \%, for size $200\times200$ 
in and the busy rate of arrival. However, it becomes negligible for larger problem sizes, even in busy scenarios.

\section{Conclusions} \label{sect:conclusions}
This work presented in this paper describes a task-based abstraction for programming FPGAs that enables task preemption using dynamic partial reconfiguration. Integrating with the existing Controller framework, we provide an approach which abstracts the low-level details around generation of a dynamic partial reconfiguration capable system and provide the programmer with a high-level API for simple management of kernel launch, data transfer and transparent book-keeping for context preemption. Furthermore, we provide transparent support for task scheduling by leveraging the default full reconfiguration mechanism that allows us to compare against the use of dynamic partial reconfiguration. 

In this paper we have demonstrated that our approach enhances flexibility by reducing the service time of urgent tasks due to the ability to dynamically swap tasks in and out. The overhead of preemptive against non-preemptive scheduling when using dynamic partial reconfiguration is 10 $\pm$ 5\% in the worst case for two reconfigurable regions, but it becomes negligible for larger problem sizes. Our experiments also demonstrate significant performance gains over the traditional use of full reconfiguration, illustrating that the use of a partial reconfiguration approach is crucial if one is to achieve low latency and flexible task scheduling on FPGAs.

For future work we plan to port our approach to the AMD Xilinx Versal architecture, which is not only a much larger FPGA which will be able to contain more complex tasks, but furthermore it will be interesting to explore how the hardened elements, such as the Network on Chip (NoC), interact with our approach. It will also be interesting to study task migration between FPGAs and othe architectures such as GPUs and CPUs, potentially moving running tasks to a more suited architecture mid-execution. Lastly we plan to explore the potential for reducing the overheads of dynamic partial reconfiguration by leveraging a custom ICAP controller, as the one provided by AMD Xilinx is only capable of exploiting up to 2.5\% of the port's bandwidth \cite{vipin2018fpga} and thus results in significantly slower reconfiguration performance.

%textcolor{red}{The source code of this work is publicly available in TBC \emph{(this will be made public in the next couple of weeks before the paper is accepted, and-so the URL will be present in the final version of the paper.)}}

\section{Acknowledgements}
The authors acknowledge EPCC at the University of Edinburgh and EPSRC who have funded this work and the ExCALIBUR H\&ES FPGA testbed for providing the FPGA compute resource. This research has been partially funded by Junta de Castilla y León - FEDER Grants, project PROPHET-2 (VA226P20).

\bibliographystyle{ACM-Reference-Format}
\bibliography{bibliography/bibliography}
%
%\subsection{Appendix Section}
%
%\begin{verbatim}
%\appendix
%
%\section{Section title of first appendix\label{app1}}
%.
%.
%.
%
%\end{verbatim}

\end{document}